\documentclass[aps,prx,twocolumn,superscriptaddress,nobalancelastpage,amsmath,amssymb,longbibliography,groupedaddress,nofootinbib]{revtex4-1}

\usepackage[utf8]{inputenc}
\usepackage{float}
\usepackage{amsmath}
\usepackage{tikz}
\usetikzlibrary{decorations.pathreplacing}
\usetikzlibrary{arrows,arrows.meta,calc,shapes.geometric,shapes.misc}
\tikzset{
	>=stealth',
	help lines/.style={dashed, thick},
	important line/.style={thick},
	connection/.style={thick, dotted},
}

\usepackage{graphicx}

\usepackage{framed, ulem}
\usepackage{bm}
\usepackage{hyperref}
\hypersetup{
    pdftitle={Z2 top order},
    unicode=false,          
    pdftoolbar=true,        
    pdfmenubar=true,        
    pdffitwindow=false,     
    pdfstartview={FitH},    
    pdfsubject={},   
    pdfcreator={},   
    pdfproducer={}, 
    pdfkeywords={} {} {}, 
    pdfnewwindow=true,      
    colorlinks=true,       
    linkcolor=blue, 
    citecolor=blue,        
    filecolor=blue,      
    urlcolor=blue           
}

\usepackage{wasysym}

\makeatletter
\newsavebox{\@brx}
\newcommand{\llangle}[1][]{\savebox{\@brx}{\(\m@th{#1\langle}\)}%
  \mathopen{\copy\@brx\kern-0.5\wd\@brx\usebox{\@brx}}}
\newcommand{\rrangle}[1][]{\savebox{\@brx}{\(\m@th{#1\rangle}\)}%
  \mathclose{\copy\@brx\kern-0.5\wd\@brx\usebox{\@brx}}}
\makeatother

\begin{document}
\title{Prediction of Toric Code Topological Order from Rydberg Blockade}

\author{Ruben Verresen}
\author{Mikhail D. Lukin}
\author{Ashvin Vishwanath}
\affiliation{Department of Physics, Harvard University, Cambridge, MA 02138, USA}

\begin{abstract}
The physical realization of $\mathbb Z_2$ topological order as encountered in the paradigmatic toric code has proven to be an elusive goal.
We predict that this phase of matter can be realized in a two-dimensional array of Rydberg atoms placed on the ruby lattice, at specific values of the Rydberg blockade radius.
First, we show that the blockade model---also known as a `PXP' model---realizes a monomer-dimer model on the kagome lattice with a single-site kinetic term. This can be interpreted as a $\mathbb Z_2$ gauge theory whose dynamics is generated by monomer fluctuations.
We obtain its phase diagram using the numerical density matrix renormalization group method and find a topological quantum liquid (TQL) as evidenced by multiple measures including (i) a continuous transition between two featureless phases, (ii) a topological entanglement entropy of $\ln 2$ as measured in various geometries, (iii) degenerate topological ground states and (iv) the expected modular matrix from ground state overlap.
Next, we show that the TQL persists upon including realistic, algebraically-decaying van der Waals interactions $V(r) \sim 1/r^6$ for a choice of lattice parameters.
Moreover, we can directly access topological loop operators, including the Fredenhagen-Marcu order parameter. We show how these
can be measured experimentally using a dynamic protocol, providing a ``smoking gun'' experimental signature of the TQL phase. Finally, we show how to trap an emergent  anyon and realize different topological boundary conditions, and we discuss the implications for exploring fault-tolerant quantum memories.
\end{abstract}

\date{\today}
\maketitle

\tableofcontents

\section{Introduction}
Nearly five decades ago, Anderson \cite{ANDERSON} proposed that quantum fluctuations could lead to a liquid of resonating valence bonds,  stimulating a vast theoretical effort that continues to this day.
Further work related  this idea to the more precise  notion of a gapped quantum spin liquid\footnote{In this work, we will interchangeably refer to this as a topological quantum liquid or spin liquid---even if the bosonic degrees of freedom are not spins but represent, e.g., a two-level atomic state.}, an exotic state  potentially realized in frustrated magnets \cite{Read91,Wen,Sachdev_Triangle,SpinLiquidsReviewBalents}. At the same time, it was understood that such gapped quantum liquids involve topological order \cite{Witten89, Wen89, Wen_90}, the simplest example being $\mathbb Z_2$ topological order in two spatial dimensions \cite{Read91,Wen}.
 
Phases of matter with topological order exhibit a number of remarkable properties \cite{Wenbook}. First, they imply the emergence of gauge fields, analogous to those describing the fundamental forces, although the gauge group and other details differ. Thus, $\mathbb Z_2$ topological order is associated with a deconfined $\mathbb Z_2$ (Ising) gauge group \cite{Wegner71,FradkinShenker}. Second, despite being built from bosonic degrees of freedom, the excitations of such quantum spin liquids are quasiparticles with nontrivial quantum statistics \cite{Wilczek82}. For example, the $\mathbb Z_2$ spin liquid includes three nontrivial excitations, two of which, the electric and magnetic particles, $e$ and $m$, are bosons, while their combination $f=em$ is a fermion \cite{Kivelson_1987,Read89,Kivelson89}. All three particles acquire a sign change on circling another anyon, i.e., they have semionic mutual statistics. These nontrivial statistics immediately lead to the remarkable property that the ground states of a topologically ordered system must be degenerate when realized on certain manifolds, such as a torus \cite{Wen89}. Third, there is a remarkable link between superconductivity and $\mathbb Z_2$ quantum spin liquids \cite{Anderson87,BASKARAN,Kivelson_1987,Baskaran88,Affleck88b,Sachdev_1991,Read89,RK,Kivelson_1987,Balents, SenthilFisher}---while the fermion $f$ can be associated with Bogoliubov quasiparticles, the $e,\,m$ excitations are related to the superconductor vortices. This led to earlier proposals suggesting that $\mathbb Z_2$ topological order might be key to understanding the phenomenon of high-temperature superconductivity.
 
Finally, a key characteristic of topological order---the long-ranged nature of its entanglement---was pointed out \cite{Kitaev_2003}. On the one hand, this implies that topologically ordered states of matter realize an entirely new form of entangled quantum matter, unlike any other conventional ground states realized to date \cite{Hamma05,Kitaev06b,Levin06}. On the other hand, this observation also has profound implications in areas such as quantum error correction and fault-tolerant quantum computation. The $\mathbb Z_2$ topological order underlies the `toric code' \cite{Kitaev_2003,Kitaev06} and `surface code' \cite{Fowler_2012} models for topologically protected quantum memory, which encode logical quantum bits in degenerate ground states. Since these degenerate ground states cannot be distinguished by local measurements, quantum information encoded in them is naturally protected from decoherence. Such intrinsic topological fault tolerance is of great consequence in the quest to build robust quantum information processing devices \cite{Nayak_RMP,Terhal_2015}. 

Due to these considerations, realizing $\mathbb Z_2$ topological order has been a major goal of condensed matter research. Unfortunately, despite several decades of theoretical and experimental effort \cite{Sachdev99,Sachdev_Triangle,Moessner01,Moessner_2001,BFG,Misguich02,Senthil02,Motrunich02,Sheng05,Kitaev06,Yan_2011,Roychowdhury15,Chamon20,Zhou20}, no clear-cut realization of $\mathbb Z_2$ topological order has been obtained to date. While topologically ordered states appear in the context of the fractional quantum Hall effect \cite{Tsui82,Laughlin83}, they are realized under rather special conditions of strong magnetic fields. In contrast, realizing topological order in a time-reversal invariant system remains a major unfulfilled research goal. Such a realization would avoid the need for applying  strong magnetic fields, which is particularly challenging for neutral objects. Furthermore, non-chiral topological orders can be achieved, in which a gap can be maintained even at the boundaries. 
In fact, we note that no realization of topological order in an intrinsically bosonic or spin system has been conclusively identified to date \cite{Broholmeaay0668}.

\begin{figure}
    \centering
    \includegraphics[scale=1]{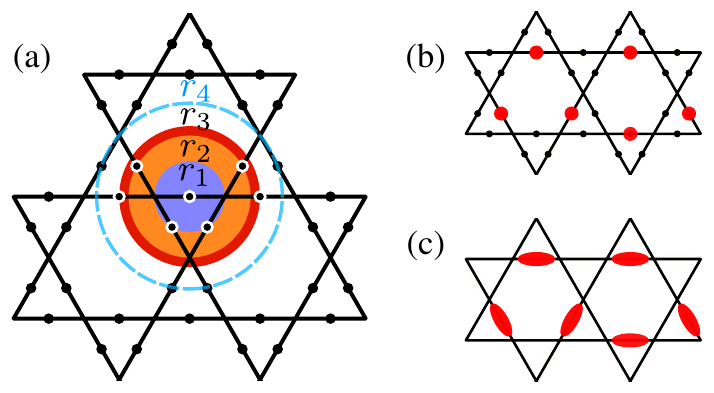}
    \caption{\textbf{Rydberg blockade model and relation to dimer model.} (a) Hardcore bosons on the links of the kagome lattice (forming the ruby lattice) are strongly-repelling, punishing double-occupation within the disk $r\leq r_3 = 2a$. (b) An example of a state consistent with the Rydberg blockade at maximal filling. (c) Since the blockade forbids occupation of any two touching bonds, we can equivalently draw the configuration as a dimer covering on the kagome lattice.}
    \label{fig:model}
\end{figure}

Recently, a new approach for exploring quantum many body physics has emerged. It is based on neutral atom arrays trapped in optical tweezer arrays. Tunable atom interactions can be engineered in such systems using the Rydberg blockade mechanism \cite{Jaksch00,Lukin01,Gaetan09,Urban09}, mediated by laser excitation of atoms into the Rydberg states \cite{Browaeys_2020,RevModPhys.82.2313}. Significant progress in realizing two dimensional quantum lattice models from the atom arrays was achieved, and a rich phase diagram of symmetry breaking orders has been predicted and observed \cite{PhysRevLett.124.103601,Ebadi20,Scholl20}. At the same time, the special features of the Rydberg atom interactions make them attractive platforms for realizing emergent lattice gauge theories and quantum dimer models \cite{Weimer10,Weimer11,PhysRevX.10.021057,Glaetzle_2014,Tagliacozzo13,Banuls20}. We note that a symmetry-protected topological phase has been realized in one-dimensional Rydberg chains \cite{deLeseleuc19}; this is distinct from the intrinsic topological order considered in this work, which does not require any symmetries and is characterized by emergent anyons.

Here we introduce a new approach for realizing a $\mathbb Z_2$ topologically ordered state as the ground state of a 2D Rydberg atom array. We show that this approach does not require careful engineering or fine-tuning of the constraints, enabling the first realization and direct probing of a time reversal and parity invariant topological order, and of emergent deconfined gauge fields in a quantum model on a near-term quantum device.

Our approach for realizing a topological spin liquid is based on the Rydberg blockade \cite{Jaksch00,Lukin01,Gaetan09,Urban09}:
when a neutral atom is excited into a Rydberg state with a high principal quantum number, the resonant excitation of the nearby atoms is suppressed due to strong atom-atom interactions. 
A minimal effective Hamiltonian for the Rydberg array---where the possibility of exciting an atom into a Rydberg state is described by two-level system---is the so-called $PXP$ model $H = \frac{1}{2} \sum_i \left( \Omega \; P \sigma^x_i P - \delta \; \sigma^z_i \right)$. Here $P$ projects out states that violate the blockade, and $\Omega$ is the Rabi frequency between the two levels, which is driven by a laser with detuning $\delta$. For a fixed blockade radius, the PXP model \cite{Sachdev02}, which depends on a single parameter $\delta/\Omega$, has been explored in great detail in one dimension---both theoretically and experimentally---where it led to a rich phenomenology including quantum scars \cite{Bernien17,Turner18a,Turner18b,Choi19,Lin_2019,Shiraishi19,Mark20,Moudgalya20,Iadecola20} and lattice gauge theories \cite{Surace20}. Recently, 2D PXP models have also been studied in the context of quantum scars \cite{Lin20,Michailidis20}. In this work, we show that for a particular choice of two-dimensional atom arrangement, Rydberg blockade radius and laser detuning, a $\mathbb Z_2$ spin liquid is stabilized as the ground state of this model.

To be specific, we first focus on the $PXP$ model on the so-called ruby lattice---equivalently, the links of the kagome lattice---with the blockade radius containing six nearby sites (see Fig.~\ref{fig:model}(a)). By tuning $\delta$, we find a phase transition from the trivial phase into another featureless phase of matter. We determine that the latter is a $\mathbb Z_2$ spin liquid using a variety of probes, including the topological entanglement entropy, ground state degeneracies, and modular transformations. We note that there are previous works considering spin liquids on the ruby lattice \cite{Bombin09,Hu11,Buerschaper14,Jahromi16,Rehn17}, although they are all distinct from the present work; in particular they feature fundamentally distinct spin interactions and do not invoke a Rydberg blockade.

These results can be understood by noting that for the above lattice and Rydberg blockade radius, the Hamiltonian becomes equivalent to a dimer-monomer model on the kagome lattice. While it is known that dimer models on non-bipartite lattices \cite{Read91,Sachdev92} (such as the triangular \cite{Moessner01} and kagome lattice \cite{Misguich02}) can realize a $\mathbb Z_2$ spin liquid, they are notoriously hard to implement in experiment. Indeed, even to realize the Hilbert space of a dimer model requires  special interactions. Furthermore, one needs the right Hamiltonian to drive the model into a spin liquid phase. For instance, Ref.~\cite{Misguich02} discovered a remarkable exactly-soluble $\mathbb Z_2$ dimer liquid on the kagome lattice, which however requires 32 distinct dimer resonances. If one only includes the lowest order dimer moves, a valence bond solid is realized \cite{Nikolic03,Singh07} rather than a spin liquid. The novel insight in the present work is that by including monomers, the effective Hamiltonian only needs a \emph{single-site} kinetic term (the creation and destruction of monomers) to perturbatively generate the multi-site dimer resonances necessary for a spin liquid. While dimer-monomer models have a rich history \cite{RK,Sylj05,Poilblanc06,Ralko07,Poilblanc10,Schwandt10}, to the best of our knowledge they have not yet been studied with a minimal kinetic term generating a rich phenomenology. Dimer-monomer models of this type could provide a new paradigm for the physical realization of lattice gauge theories, going well beyond the example studied in this work.

Furthermore, we show that the above findings are not fine-tuned to the $PXP$ model. More precisely, we numerically confirm that the spin liquid can also be found in the full-fledged Hamiltonian with realistic $V(r) \sim 1/r^6$ Van der Waals interactions between the Rydberg atoms on a particular instance of the ruby lattice.

In addition to realizing a $\mathbb Z_2$ spin liquid in an experimentally-relevant model, a very useful property of this model is that it also gives a direct handle on the two topological string operators. In the language of lattice gauge theory, these are the Wilson and 't Hooft lines. In the context of topological order, these are the strings whose endpoints host an $e$- and $m$-anyon, respectively. We explicitly construct these operators on the lattice and confirm the expected behavior of loop operators in the spin liquid, as well as re-interpret the nearby phases as $e$- and $m$-condensates using the Fredenhagen-Marcu string order parameter 
\cite{Fredenhagen83,Fredenhagen86,Marcu86,Fredenhagen88,Gregor11}.

These string operators also serve as very useful probes to detect the spin liquid in experiments.
The possibility of measuring nonlocal observables is truly a remarkable advantage of certain cold-atom platforms \cite{Endres11,Hilker17,Chiu19,deLeseleuc19}. In more conventional solid state systems, one must rely on local probes which are suited to identifying local order parameters but cannot directly detect topological order. In contrast, Rydberg platforms allow one to take snapshots of the quantum state with single-site resolution, opening up the possibility of extracting nonlocal correlation functions. We describe in detail how this feature can be deployed to diagnose topological order. While the diagonal string operator can be readily measured, we further show how the string operator for the $e$-anyon---which a priori involves off-diagonal operations which are hard to measure in the lab---can be converted into a diagonal string operator by time-evolving with a Hamiltonian whose blockade radius has been quenched. Thus, we show that \emph{both} string operators become measurable in the diagonal basis

Finally, we discuss methods to create and manipulate quantum information stored in topologically degenerate ground states, paving the way for potential exploration of topological quantum memories. Two crucial pieces of the puzzle we identify are the ability to trap an $e$-anyon and to create distinct topological boundary conditions---both are straightforwardly achieved by locally changing the laser detuning. As we will explain, these two ingredients already give access to topologically-degenerate qubits in the plane which can be initialized and read out.

The remainder of the paper is structured as follows. Section~\ref{sec:blockade} concerns the Rydberg blockade model, with subsection~\ref{sec:dimer} comparing it to and distinguishing it from conventional dimer models. Its phase diagram is obtained in subsection~\ref{sec:phasediagram}, containing a trivial phase, a $\mathbb Z_2$ spin liquid, and a valence bond solid. We confirm that the intermediate phase is indeed a spin liquid in terms of its topological entanglement entropy (subsection~\ref{sec:topo}), its topological string operators (subsection~\ref{sec:strings}) and its topologically-distinct ground states from which we extract part of the modular matrices (subsection~\ref{sec:gs}). Section~\ref{sec:realization} focuses on the experimental feasibility, with subsection~\ref{sec:vanderwaals} showing that the spin liquid persists upon including the $V(r)\sim 1/r^6$ potential and subsection~\ref{sec:offdiagonal} explaining how the off-diagonal string operator can be reduced to a diagonal observable. We end with section~\ref{sec:qubit} taking the first steps towards using this novel realization for creating a fault-tolerant quantum memory by showing how to trap $e$-anyons (subsection~\ref{sec:e}) and how to realize distinct boundary conditions (subsection~\ref{sec:boundary}); section~\ref{sec:surface} then gives examples of how this can be applied.

\section{Rydberg blockade `PXP' model \label{sec:blockade}} 

We consider hardcore bosons on the links of the kagome lattice with a two-dimensional version of the Fendley-Sengupta-Sachdev model \cite{Sachdev02,Fendley04}:
\begin{equation}
H = \frac{\Omega}{2} \sum_{\bm i} \left( b_{\bm i}^{\vphantom \dagger} +  b_{\bm i}^\dagger \right) - \delta \sum_{\bm i} n_{\bm i} + \frac{1}{2}\sum_{\bm i,\bm j} V(|\bm i - \bm j|) \; n_{\bm i} n_{\bm j}.  \label{eq:ham}
\end{equation}
We set\footnote{Note that the sign of $\Omega$ can be toggled by replacing $b_i \to -b_i$, which leaves $n_i$ invariant. The only place in this paper where the sign of $\Omega$ matters is in the definition of the topological string operators; see section~\ref{sec:strings} .} $\Omega>0$.
For Rydberg atoms, $V(r) \sim 1/r^6$. We defer that case to section~\ref{sec:realization}.
Here, we instead focus on the simpler model where $V(r)$
forms a blockade in a particular disk:
\begin{equation}
V(r) = \left\{ \begin{array}{lll}
+\infty & & \textrm{ if } r \leq 2a \\
0 & & \textrm{ if } r > 2a.
\end{array} \right. \label{eq:V}
\end{equation}
Here the lattice spacing $a$ is the shortest distance between two atoms. As shown in Fig.~\ref{fig:model}(a), with this interaction range, a given site is coupled to six other sites, which are ordered in pairs at distances $r_1 = a$, $r_2=\sqrt{3} a \approx 1.73a$ and $r_3 = 2a$ (the next distance would be $r_4 = \sqrt{7} a \approx 2.65a$, denoted by the dashed circle in Fig.~\ref{fig:model}(a)).
The Rydberg blockade implies that any two sites within this distance cannot both be occupied (Fig.~\ref{fig:model}(b)), which we can interpret as a dimer state on the kagome lattice if the system is at maximal filling (see Fig.~\ref{fig:model}(c)).
We note that this blockade Hamiltonian is equivalent to the PXP model mentioned in the introduction.

\subsection{Connection to and differences from dimer models \label{sec:dimer}}

For a dimer state on the kagome lattice, each vertex is touched by exactly one dimer, such that $\langle n \rangle = \frac{1}{4}$. Our model can have $\langle n \rangle < \frac{1}{4}$, in which case certain vertices have \emph{no} dimers---referred to as a \emph{monomer}. This distinguishes our system from a usual dimer model. Let us briefly discuss the implications of this difference. The reader interested in the numerical results for our model can skip ahead to section~\ref{sec:phasediagram}.

The constraint of a dimer model---having exactly \emph{one} dimer per vertex---can be interpreted as a Gauss law \cite{Fradkin90}. More precisely, the presence or absence of a dimer represents a $\mathbb Z_2$-valued electric field, with the dimer constraint enforcing the lattice version of the Gauss law $\bm{\nabla \cdot E} = 1 \,\rm ( mod \, 2)$.
Each vertex thus carries a classical/static electric charge $e$. For this reason, a dimer model is also referred to as an odd $\mathbb Z_2$ gauge theory \cite{JalabertSachdev,Moessner_2001}.
The absence of dynamic matter in a dimer model implies that it is a pure $\mathbb Z_2$ gauge theory, which has two possible phases: a deconfined\footnote{This refers to the freedom of test charges which in this case are monomers.} and a confined phase. The former is our desired $\mathbb Z_2$ spin\footnote{A more apt name would be a dimer liquid, but here we follow the more common terminology ingrained in the literature.} liquid (or equivalently, a resonating valence bond state), whereas the latter is a valence bond solid\footnote{The confined phase is a condensate of the magnetic excitation $m$. As explained in section~\ref{sec:strings}, this anyon carries a projective representation under translation such that its condensation implies translation symmetry breaking.}. Stabilizing the spin liquid requires dimer resonances in the Hamiltonian, but due to the local constraint of a dimer model, these terms typically span many sites. The smallest resonance acts on the six sites around a hexagon of the kagome lattice. The solvable dimer model by Misguich, Serban and Pasquier \cite{Misguich02} requires 32 distinct types of resonances, the largest spanning 12 sites. While these conditions can be somewhat relaxed \cite{Hao_2014}, the direct implementation of dimer models, tuned to a  regime of parameter space where a liquid phase is known to emerge,  remains extremely challenging.

In contrast, the Rydberg blockade model \eqref{eq:ham} is a dimer-monomer model. In other words, the Gauss law of the lattice gauge theory is now $\bm{\nabla \cdot E} = \rho$, where $\rho$ is a quantum-mechanical two-level degree of freedom. This has two advantages. Firstly, the only explicit dynamics in our model is a \emph{single-site} term which creates and destroys pairs of monomers/charges (the Rabi oscillation $\Omega$ in Eq.~\eqref{eq:ham}). In the limit of large $\frac{\delta}{\Omega}$, the low-energy theory is projected into the macroscopically degenerate space of (maximally-filled) dimer states. Virtual monomer excitations induce dimer resonances between these states. For instance, at leading order in perturbation theory, we obtain $H_{\textrm{eff}} = - \frac{3\Omega^6}{32\delta^5} \sum_{\hexagon} \Big( \big |\begin{tikzpicture}[baseline=-0.65ex,scale=0.23]
\def\a{0.5};
\def\b{0.866};
\draw (1,0) -- (\a,\b) -- (-\a,\b) -- (-1,0) -- (-\a,-\b) -- (\a,-\b) -- (1,0);
\filldraw[rotate around={60:(-0.75,\b/2)},color=red] (-0.75,\b/2) ellipse (0.4 and 0.15);
\filldraw[rotate around={-60:(0.75,\b/2)},color=red] (0.75,\b/2) ellipse (0.4 and 0.15);
\filldraw[color=red] (0,-\b) ellipse (0.4 and 0.15);
\end{tikzpicture}
\big \rangle
\big \langle
\begin{tikzpicture}[baseline=-0.65ex,scale=0.23]
\def\a{0.5};
\def\b{0.866};
\draw (1,0) -- (\a,\b) -- (-\a,\b) -- (-1,0) -- (-\a,-\b) -- (\a,-\b) -- (1,0);
\filldraw[rotate around={60:(0.75,-\b/2)},color=red] (0.75,-\b/2) ellipse (0.4 and 0.15);
\filldraw[rotate around={-60:(-0.75,-\b/2)},color=red] (-0.75,-\b/2) ellipse (0.4 and 0.15);
\filldraw[color=red] (0,\b) ellipse (0.4 and 0.15);
\end{tikzpicture}
\big | + h.c. \Big)$, describing hexagon resonances.
Second, since monomers are now dynamical degrees of freedom, they can be condensed, driving the system to a translation-symmetric trivial state\footnote{In the language of $\mathbb Z_2$ gauge theory coupled to matter, this corresponds to the Higgs phase.}. This gives a clear-cut instance of a continuous phase transition between two featureless phases of matter (as opposed to the valence bond solid, which has long-range order), which does not involve any symmetries.

While there are thus clear advantages to not realizing a strict dimer model but rather a dimer-monomer model, it is also advantageous to nevertheless be proximate to a dimer model (i.e., have low monomer density). Firstly, it is a good place to hunt for a spin liquid, since---as discussed above---a dimer model on the kagome lattice cannot realize a trivial phase of matter. Secondly, one has a direct handle on the topological string operators associated to the $\mathbb Z_2$ gauge theory, with anyons living at their endpoints. We discuss this in detail in section~\ref{sec:strings}.

\begin{figure*}
	\centering
	\includegraphics[scale=0.32]{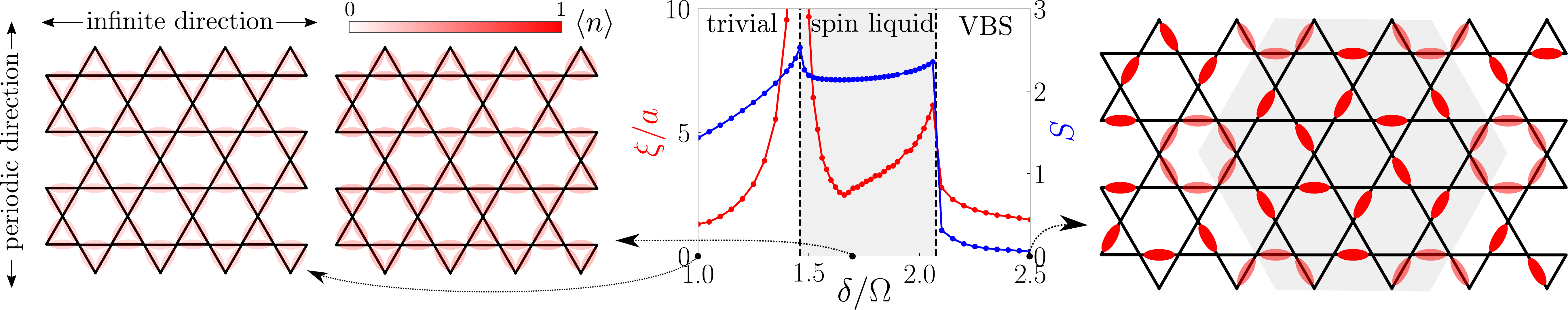}
	\caption{\textbf{Phase diagram of Rydberg blockade model on the links of the kagome lattice.} The trivial phase at small $\delta/\Omega$ is separated from the valence bond solid (VBS) at large $\delta/\Omega$ by an intermediate phase which has a large entanglement plateau. We show an exemplary density plot for each of the three phases, which shows that the intermediate phase is featureless. The VBS phase has a 36 site unit cell (72 atoms on the links) highlighted by the gray shaded region---this pattern was studied before in Refs.~\cite{Nikolic03,Singh07,Poilblanc11} in the context of the spin-$1/2$ Heisenberg model on the kagome lattice. Numerical results are for a cylinder with XC-8 geometry, as depicted.}
	\label{fig:phasediagram}
\end{figure*}

\subsection{Phase diagram \label{sec:phasediagram}}

We now study the phase diagram of the model in Eq.~\eqref{eq:ham} with the blockade in Eq.~\eqref{eq:V} using the density matrix renormalization group (DMRG) \cite{White92,White93,Stoudenmire12,Hauschild18}. We can explicitly enforce $V(r_1) = +\infty$ by working in the reduced Hilbert space where each triangle of the kagome lattice (containing three atoms) only has four states: empty or a dimer on one of the three legs. We cannot straightforwardly set $V(r_2) = V(r_3) = +\infty$ since the resulting Hilbert space is no longer a tensor product
\footnote{For a review on tensor network methods in constrained systems, see Ref.~\cite{Banuls20}. The Rydberg blockade has been explicitly enforced in DMRG for 1D systems by Chepiga and Mila \cite{Chepiga19}; it would be interesting to generalize this to the current two-dimensional setting.}---indeed, this is the magic of dimer models. Hence, we enforce these constraints energetically by choosing a very large $V(r_2) = V(r_3) = 50 \Omega$. We have confirmed that our results do not depend on the details of this choice. We study the model on a cylinder geometry of fixed circumference (up to XC-12) and infinite extent \cite{Stoudenmire13}. See Appendix~\ref{app:numeric} for details about the numerical method.

When $\delta/\Omega$ is low enough, the system is adiabatically connected to the empty state and is thereby completely trivial. For very large $\delta/\Omega$ we enter the regime that is perturbatively described by a dimer model, as explained in section~\ref{sec:dimer}. We find that its ground state spontaneously breaks crystalline symmetries and forms a valence bond solid (VBS). Remarkably, for intermediate $\delta/\Omega$, these two phases are separated by another featureless phase, as shown in Fig.~\ref{fig:phasediagram} by the diverging correlation length $\xi$ and the entanglement entropy $S$ between two rings of the cylinder. We will argue that this is a $\mathbb Z_2$ spin liquid.

\begin{figure}[t]
	\centering
	\includegraphics[scale=1]{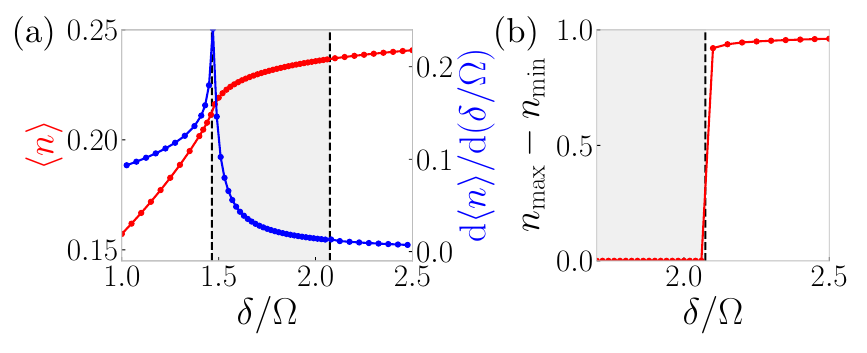}
	\caption{\textbf{Detecting phase transitions via filling fraction.} This data is obtained for an infinitely-long cylinder with XC-8 geometry. (a) The filling fraction has a singular behavior upon transitioning from the trivial phase into the spin liquid, after which the system enters a regime where $\langle n \rangle \approx 0.25$, consistent with it being an approximate dimer state.
		Note that the first derivative of $\langle n \rangle$ corresponds to a second derivative of the energy; the observed singularity is thus a sign of a second-order phase transition between the trivial phase and the spin liquid. 
		(b) The spin liquid and VBS phase are separated by a first order transition.    }
	\label{fig:n_and_P}
\end{figure}

As a first indication that this intermediate phase is still within the approximate dimer model, we consider the filling fraction $\langle n \rangle$, shown by the red curve in Fig.~\ref{fig:n_and_P}(a). We see that as $\delta/\Omega \to \infty$, the filling $\langle n \rangle$ approaches the maximal $1/4$ consistent with a fully packed dimer picture. In the intermediate regime (shaded in the plot) we are no longer in the VBS phase, but $\langle n \rangle$ is still large. It is only when $\delta/\Omega$ is decreased further---entering the trivial phase---that $\langle n \rangle$ sharply drops. This is in line with the possible scenario of exiting the spin liquid by condensing monomers---as explained in section~\ref{sec:dimer}---which would exhibit itself in a rapid drop of filling density.

Moreover, the derivative of $\langle n \rangle$ diverges at the transition between the trivial phase and the spin liquid, signaling a continuous transition.
Indeed, the theoretical expectation is that this belongs to the $2+1D$ Ising universality class (with the trivial phase corresponding to the `ordered' side), but our available system sizes are not big enough to accurately extract scaling dimensions. Fig.~\ref{fig:n_and_P}(a) shows no such singularity between the spin liquid and VBS phase. However, it turns out that it is a first order transition which is very hard to diagnose this way (due to the small energy scales associated to the VBS phase). This is much more easily demonstrated by considering the variation of $\langle n \rangle$ between different sites: Fig.~\ref{fig:n_and_P}(b) shows that this jumps discontinuously.

\subsection{Topological entanglement entropy \label{sec:topo}}

One characteristic feature of topological phases of matter can be found in the scaling of the entanglement entropy. Gapped phases of matter satisfy an area law: for a region with perimeter $L$, we have $S(L) = \alpha L - \gamma$. The constant offset $\gamma$ is a universal property called the topological entanglement entropy, encoding information about the quantum dimensions of the anyons of the topological order \cite{Kitaev06b,Levin06}. For a $\mathbb Z_2$ spin liquid, $\gamma = \ln 2$ \cite{Hamma05}.

\begin{figure}
	\centering
	\includegraphics[scale=1]{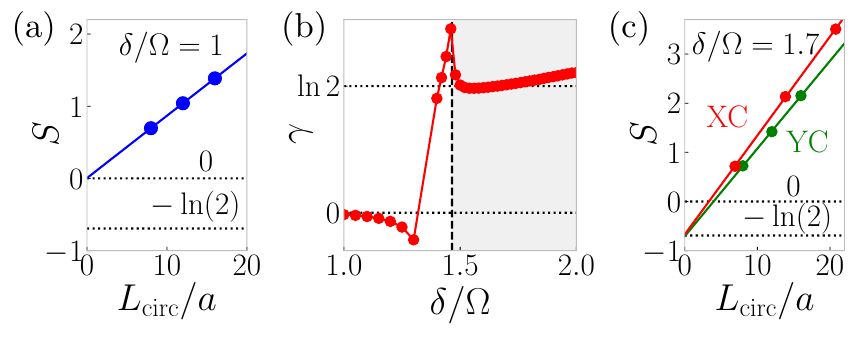}
	\caption{\textbf{Topological entanglement entropy.} We determine the offset $\gamma$ in the area law $S = \alpha L - \gamma$ \cite{Kitaev06b,Levin06,Jiang12,Jiang13}. (a) For the trivial phase, this is zero. (b) As we increase $\delta/\Omega$, we enter the spin liquid where $\gamma \approx \ln 2$. Here, we plot $S_{L=8} - 2S_{L=4}$ where $S_{L=n}$ is the bipartition entanglement entropy for the XC-$n$ geometry. (c) For an exemplary point in the spin liquid, we extract $\gamma$ for two distinct geometries (up to XC-12 and YC-8). Note that XC-$n$ (YC-$n$) has circumference $L_\textrm{circ}/a = \sqrt{3}n$ ($2n$).}
	\label{fig:Stopo}
\end{figure}

The topological entanglement entropy can be efficiently extracted from a cylinder geometry \cite{Jiang12,Jiang13}. We take a point in the middle of the presumed spin liquid in Fig.~\ref{fig:phasediagram}, $\delta/\Omega= 1.7$, and numerically obtain the entanglement entropy upon bipartitioning the infinitely-long cylinder in two halves. Doing this for different circumferences\footnote{As will be explained in section~\ref{sec:gs}, there are degenerate ground states on the cylinder. To make sure we are comparing apples to apples, we choose the {$|1\rangle$} ground state on each cylinder as determined by the string operators discussed in section~\ref{sec:strings}.}, we extract $\gamma \approx \ln 2$, as shown in Fig.~\ref{fig:Stopo}(c).
Importantly, it has been observed before that one can obtain spurious value of $\gamma$ for specific cuts in certain lattice models, i.e., one can be deceived into thinking a trivial phase is in fact topologically ordered \cite{Gong13,Gong14,Zou16,Williamson19}. For all such reported cases, the spurious value can be detected by comparing the results for different cuts\footnote{E.g., for the $J_1$-$J_2$ model on the honeycomb lattice, Fig.~14(b) of Ref.~\cite{Gong13} shows that the AC geometry gives $\gamma \approx \ln2$ whereas the tZC geometry gives $\gamma \approx 0$. For similar results on the square lattice, see Fig.~12 of the Supplemental Materials of Ref.~\cite{Gong14} (or Fig.~17 of the arXiv version).}.
For this reason, we have extracted $\gamma$ for two distinct geometries: XC (where the finite periodic direction bisects triangles of the kagome lattice) and YC (where the circumference runs parallel to one of the axes of the kagome lattice); for an explanation of this naming convention, see Appendix~\ref{app:numeric}. Both linear fits give a topological entanglement entropy which is remarkably close to $\ln 2$. For comparison, for a point in the trivial phase ($\delta/\Omega = 1$) we obtain $\gamma=0$ (Fig.~\ref{fig:Stopo}(a)).

To confirm that the above is not a fine-tuned feature of a particular point in the phase diagram, we extract $\gamma$ as a function of $\delta/\Omega$. Figure~\ref{fig:Stopo}(b) indeed shows a plateau where $\gamma \approx \ln 2$, consistent with a $\mathbb Z_2$ spin liquid. Since this plot only relies on XC-$4$ and XC-$8$ data, there is still some minor variation within this plateau. Deep in the spin liquid, $\delta/\Omega=1.7$, we were also able to converge to the ground state on the bigger cylinder XC-$12$, confirming $\gamma \approx \ln 2$ (Fig.~\ref{fig:Stopo}(c)). Note that we do not consider $\gamma$ in the VBS phase since due to the large unit cell (shown in Fig.~\ref{fig:phasediagram}) the next consistent geometry\footnote{There is also a VBS phase on, e.g., XC-4 and XC-12, but they have different patterns and it is thus not meaningful to compare their entropies.} is XC-16, which is out of reach with current methods.

\begin{figure*}
	\centering
	\includegraphics[scale=1]{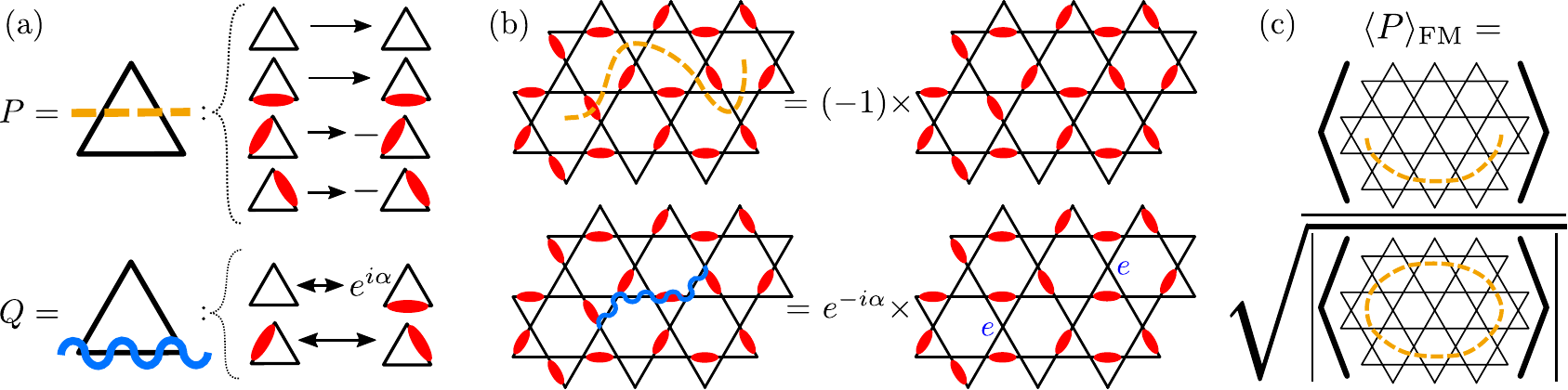}
	\caption{\textbf{Topological string operators.} (a) The two different string operators are defined by their action on a single triangle. We call the diagonal and off-diagonal string operators $P$ and $Q$, respectively. (b) An example of the action of the string operators on a classical dimer state. (c) The definition of the Fredenhagen-Marcu order parameter
		\cite{Fredenhagen83,Fredenhagen86,Marcu86,Fredenhagen88,Gregor11}
		is shown for the diagonal string, $\langle P \rangle_\textrm{FM}$, which measures the condensation of the $m$-anyon. The analogous definition for $\langle Q \rangle_\textrm{FM}$ (not shown) measures an $e$-condensate.}
	\label{fig:strings}
\end{figure*}

\subsection{String operators and anyon condensation \label{sec:strings}}

The advantage of measuring topological entanglement entropy is that it is well-defined for any model even in the absence of microscopic identification of operators corresponding to emergent gauge theory. However, in our Rydberg blockade model, a more microscopic understanding of the spin liquid is available. Here, we can identify the topological string operators associated with this $\mathbb Z_2$ lattice gauge theory, similar to the toric code model \cite{Kitaev_2003}. Such an explicit representation of a topological quantum liquid has a variety of uses: in identifying the spin liquid and its nearby phases (especially in an experimental set-up where, e.g., topological entanglement entropy is not readily accessible), in creating anyons,  in distinguishing topological ground states and also perhaps  for  quantum information applications, such as  the initialization and read-out of topological  qubits.

A $\mathbb Z_2$ lattice gauge theory comes with two string operators determined by the electric field $E$ (defined modulo 2) and its conjugate variable, the gauge field $A$. These strings are the 't Hooft line $e^{i\pi \int E}$ and the Wilson line $e^{i\int A}$, which anticommute at intersection points. As already mentioned in section~\ref{sec:dimer}, the binary-valued electric field corresponds to a dimer configuration, with the hardcore dimer constraint acting as a Gauss law. The string operator $e^{i\pi\int E}$ thus corresponds to the parity of dimers along a string. To be precise, we define its action on a single triangle in Fig.~\ref{fig:strings}(a) (orange dashed line); we refer to this diagonal parity string as $P$. For an explicit matrix representation, see Appendix~\ref{app:rotate}. Due to the Gauss law, evaluating it along any closed loop---which has to run perpendicular to the bonds of the kagome lattice---measures the charge inside of it. In the absence of monomers---i.e., gauge charge excitations---this is simply $(-1)^\textrm{\# vertices enclosed}$ for a contractible loop, as expected of an odd Z$_2$ gauge theory. In contrast, non-contractible loops distinguish topologically-distinct sectors of the dimer Hilbert space (since this value cannot be changed by any local operator).

In the dimer basis, the dual string $e^{i\int A}$ has to be off-diagonal, shuffling the dimers. There is essentially a unique way of defining such a string that has a well-defined action on single triangles, as shown in Fig.~\ref{fig:strings}(a) (solid blue line); we refer to this string as $Q$. An example is shown in Fig.~\ref{fig:strings}(b). Note that any closed string that runs parallel to the bonds of the kagome lattice indeed maps a valid dimer configuration to another valid dimer configuration, and it is also easy to see that this string $Q$ anticommutes with $P$ whenever the strings intersect. To the best of our knowledge, this definition of the $Q$ string is novel; for dimer models, one often considers the more restrictive strings that have to pass through an alternating series of empty and filled bonds. The advantage of this more general definition is twofold: (1) it is also well-defined for states that contain monomers, and (2) with the definitions for $P$ and $Q$ in Fig.~\ref{fig:strings}(a), there is in fact a duality transformation that interchanges them, as discussed in section~\ref{sec:offdiagonal}.

The electric $e$ and magnetic $m$ excitations of this $\mathbb Z_2$ lattice gauge theory live at the endpoints of the $Q$ and $P$ strings, respectively. For instance, Fig.~\ref{fig:strings}(b) shows\footnote{Similarly, the open string $P$ in Fig.~\ref{fig:strings}(b) creates $m$ excitations, but this is hard to see since it is acting on a classical dimer state, which is an $m$-condensate.} how an open $Q$ string indeed creates a monomer at each end. These $e$ and $m$ excitations are topological since they can only be created in pairs. Moreover, whilst they are individually bosonic, the anticommuting property of the $P$ and $Q$ string encodes the fact that $e$ and $m$ have non-trivial mutual statistics; equivalently, the endpoint of the product string $PQ$ carries an emergent fermion $f$. 

The spin liquid is defined by the deconfinement of these excitations. The nearby phases correspond to condensing either the $e$ or the $m$, which respectively confines $m$ or $e$ due to the mutual statistics. Historically, the $e$-condensate is called the Higgs phase, whereas the $m$-condensate is called the confined phase (due to the charged $e$ excitations becoming confined). In an odd gauge theory, with nonzero background gauge charge at each lattice site, the latter in fact implies spontaneous symmetry breaking (i.e., a valence bond solid). The reason for this is that the $m$-anyon carries a projective representation\footnote{This is simply a restatement of the Gauss law that the parity along a loop surrounding a vertex is $-1$: this parity loop can be interpreted as the anticommutator $T_x T_y T_x^{-1} T_y^{-1}$ for the action of translations $T_{x,y}$ on the endpoint of a parity string $P$, i.e., the $m$-anyon.} under the $\mathbb Z \times \mathbb Z$ translation symmetry.

These condensates can be diagnosed by the open $P$ or $Q$ strings attaining long-range order. To properly define what this means, it is important to normalize these string operators. Indeed, generically these strings will decay to zero since the ground state has virtual $e$ and $m$ fluctuations. For this reason, Fredenhagen and Marcu \cite{Fredenhagen83,Fredenhagen86,Fredenhagen88} introduced the normalized string operator in Fig.~\ref{fig:strings}(c), which we will refer to as the FM string order parameter. This was also more recently imported into the condensed matter context---where lattice gauge theories are emergent---by Gregor, Huse, Moessner and Sondhi \cite{Gregor11}. These two string order parameters are a very useful tool for diagnosing the different phases of a lattice gauge theory: although confinement in pure gauge theories can be probed by an area law \cite{Wegner71,Wilson74}, in the presence of dynamic matter (as we have in our model) loop operators typically scale with a perimeter law \cite{FradkinShenker,Seiler82}.

\begin{figure}
	\centering
	\includegraphics[scale=0.29]{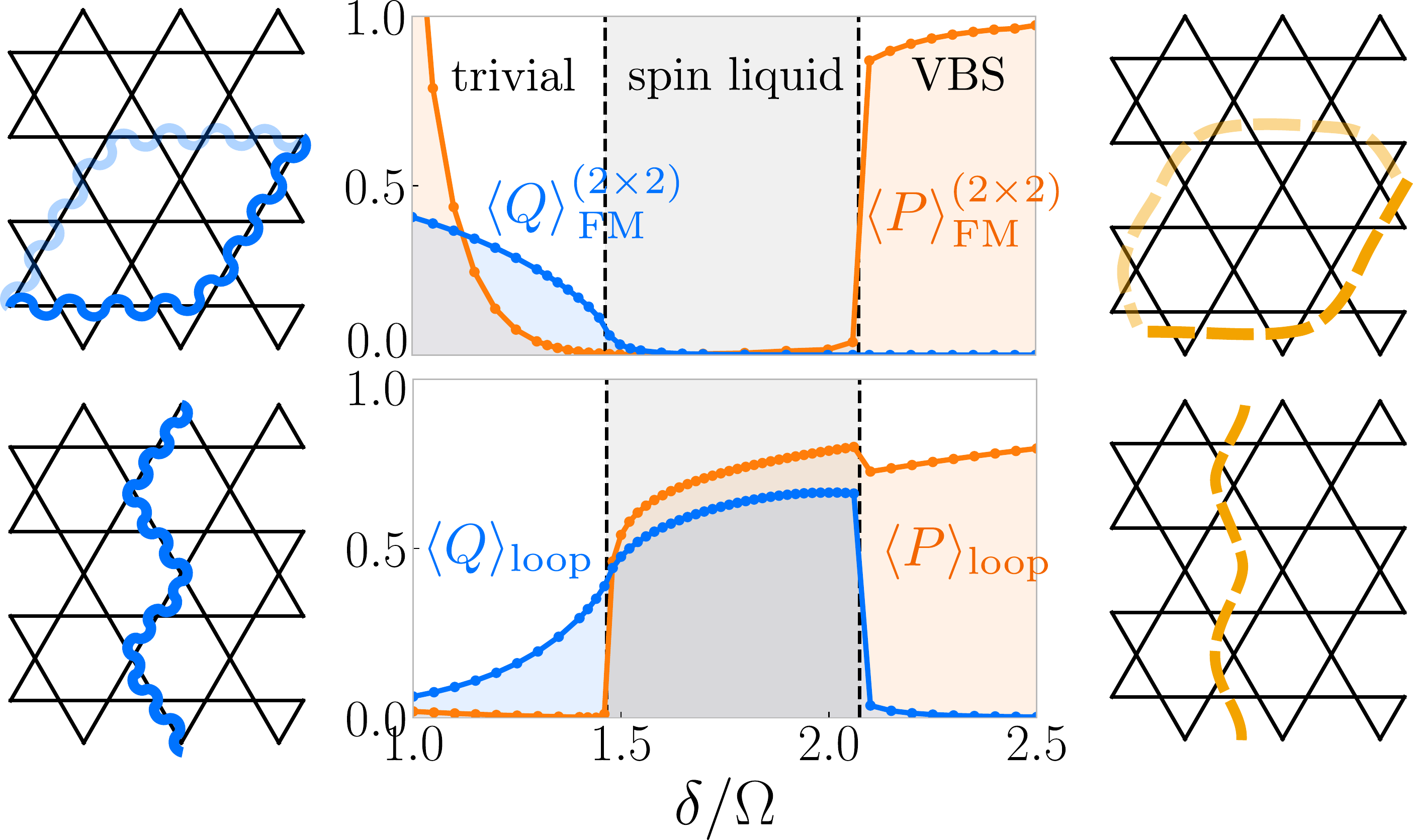}
	\caption{\textbf{Diagnosing phases in terms of topological string operators.} Top panel: the Fredenhagen-Marcu (FM) string order parameters show that the trivial phase is an $e$-condensate (= Higgs phase) and the VBS phase is an $m$-condensate (= confined phase). These string order parameters decay to zero in the spin liquid, confirming that it is the deconfined phase of the $\mathbb Z_2$ lattice gauge theory. We sketch the strings used for calculating the FM order parameter (the transparent strings show the closed loop used to normalize the string; see Fig.~\ref{fig:strings}(c)).
		Bottom panel: long-range order in the FM string for $P$ ($Q$) suppresses the value of a closed $Q$ ($P$) loop around the circumference. In the spin liquid, both loops are nonzero (we plot the absolute value: their signs label degenerate ground states, see Fig.~\ref{fig:gs}). As in Fig.~\ref{fig:phasediagram}, results are for XC-$8$ (depicted) with the vertical (horizontal) direction being periodic (infinite).}
	\label{fig:phasediagram_with_strings}
\end{figure}

The only remaining technicality to discuss is the phase factor $e^{i \alpha}$ in the definition of the off-diagonal string $Q$ in Fig.~\ref{fig:strings}(a). In general this phase factor cancels out unless the $Q$ string changes the total number of dimers, such as for the open string in Fig.~\ref{fig:strings}(b). Hence, the optimal choice of $e^{i \alpha}$ depends on the phase difference between different branches of the ground state wave function with distinct particle number. In the present model, one can straightforwardly argue that if the Rabi frequency $\Omega < 0$, then all amplitudes of the wave function have the same sign, whereas for $\Omega >0$ it alternates with the parity of dimers. From now on, we thus fix $e^{i\alpha} = - \frac{\Omega}{|\Omega|}$.

We are now in a position to evaluate the open string and loop operators in the Rydberg blockade model. The results are shown in Fig.~\ref{fig:phasediagram_with_strings}. As expected, we see that $Q$ has long-range order in the trivial phase---corresponding to an $e$-condensate---whereas $P$ has long-range order in the VBS phase---corresponding to an $m$-condensate. Note that $P$ also attains long-range order deep in the trivial phase: this is allowed since the definition of (and distinction between) $e$ and $m$ anyons is only strictly meaningful in the deconfined phase \cite{FradkinShenker}. In the intermediate spin liquid, both FM order parameters decay to zero, consistent with the claim that this is the deconfined phase of the lattice gauge theory. While Fig.~\ref{fig:phasediagram_with_strings} shows the FM string order for only particular length of the string (as depicted on both sides of the panel), a more careful scaling analysis in Appendix~\ref{app:scaling} confirms that in the spin liquid, these strings decay to zero exponentially in the length of the string. We stress that this is a very nontrivial property that would be exceedingly difficult\footnote{Indeed, due to the normalization of the strings, the only contribution can come from the endpoints, which naively only affect a finite region due to the finite correlation length. Generically, in the absence of additional symmetry properties, one expects the expectation value of operators with finite support to be nonzero. It is the emergent $1$-form symmetry of the topologically ordered phase that constrains it to be zero (up to exponentially small corrections which couple the two endpoints).} to explain without the presence of topological order. Correspondingly, in this regime, the loop operators evaluated around the circumference are not suppressed and have an appreciable value (which albeit decreases with circumference). In fact, the sign of this nonzero number labels topologically-distinct ground states, as we discuss next.

\begin{figure}
    \centering
    \includegraphics[scale=1]{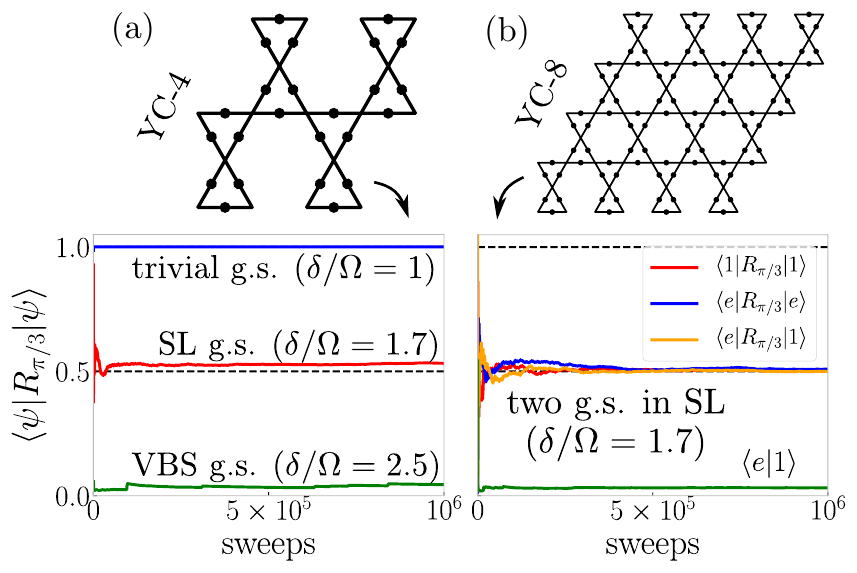}
    \caption{\textbf{Ground states and modular transformations.} From the ground states on the infinitely-long cylinder, we can obtain minimally-entangled ground states on the torus geometry. For the smaller geometry, we show that whilst the $\pi/3$-rotation acts trivially on the trivial ($\delta/\Omega=1$) or symmetry-breaking ($\delta/\Omega=2.5$) phases, it leads to a non-trivial overlap in the spin liquid ($\delta/\Omega=1.7$). We confirm for the larger torus (96 sites) that the overlaps for distinct ground states agree with the prediction \eqref{eq:overlaps} based on the modular transformation of a $\mathbb Z_2$ spin liquid. The overlaps are shown as a function of the Monte Carlo sweeps, converging toward the value $\approx 0.5$.}
    \label{fig:overlaps}
\end{figure}

\subsection{Topological ground state degeneracy and modular matrices \label{sec:gs}}

Another fingerprint of a topological spin liquid is its topological ground state degeneracy on manifolds which are themselves topologically non-trivial \cite{Wen89,Wang15,hung_ground-state_2015}. For Abelian topological order on an infinitely-long cylinder, one has a ground state corresponding to each anyon in the theory. Conceptually, these different states can be related by starting with one of the ground states and nucleating an anyon pair and separating them infinitely far along the infinite direction of the cylinder\footnote{One could of course instead choose to wrap them around the finite direction, which would generate a different basis in this four-dimensional space of states. However, these states will not be minimally-entangled on the cylinder, whereas DMRG optimizes for that \cite{Jiang12,Zhang12,Cincio13}.} For the present case, we thus expect four distinct topological ground states, corresponding to $1$, $e$, $m$ and $f$ lines threaded along the infinite axis. Due to the mutual statistics, these distinct ground states can be diagnosed by measuring the $P$ and $Q$ loops around the circumference.

Numerically, when we repeat DMRG with different random initializations, we find two (quasi-)degenerate ground states which are distinguished by the sign of $\langle P \rangle_\textrm{loop}$ around the circumference\footnote{Equivalently, if one creates an initial state with a given sign of the parity loop, we find that DMRG remains in this sector. This does not work for the $Q$ loop, presumably because its finite-size effects along the circumference are big enough for DMRG to switch sectors.}. It is tempting to associate these to the trivial anyon and the electric charge, $1$ and $e$.

\begin{figure}[t]
    \centering
    \includegraphics[scale=1]{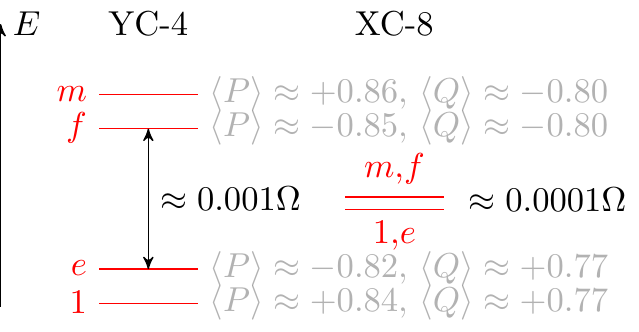}
    \caption{\textbf{Topological ground state degeneracy.} In the topological phase, we obtain the $1$ and $e$ ground states from DMRG with random initial states. The $m$ and $f$ states are obtained by starting from fixed-point resonating dimer states and subsequently applying imaginary time-evolution. The energies shown are for $\delta/\Omega=1.7$. In light gray we also show the eigenvalues of the $P$ and $Q$ loop operators around the circumference (for YC-4); the four ground states are characterized by the signs of these numbers.
    \label{fig:gs}}
\end{figure}

To make this concrete, we use the technique of Refs.~\cite{Zhang12,Cincio13}: making the resulting matrix product states (MPS) periodic along the second direction, one obtains wavefunctions on a torus geometry as shown in Fig.~\ref{fig:overlaps}, which we denote by $|1\rangle$ and $|e\rangle$. It can be shown that a $\pi/3$-rotation mixes the topological ground states. Indeed, evaluating the overlap $\langle 1| R_{\pi/3} |1 \rangle$ using quantum Monte Carlo \cite{Sandvik07}, we see for the smaller torus (of 24 sites) in Fig.~\ref{fig:overlaps}(a) that (i) the ground state of the trivial phase is completely symmetric, (ii) the ground state of the symmetry-broken phase gives a vanishing overlap with the rotated wavefunction, and (iii) the ground state in the presumed spin liquid gives a \emph{finite} overlap, suggesting that it has overlap with a finite number of other states. In fact, its value is universal and can be derived as in Ref.~\cite{Zhang12}. In particular, the relevant $2\times 2$-block of the modular matrix is
\begin{equation}
 \left( \begin{array}{cc}
\langle 1| R_{\pi/3} | 1 \rangle & \langle 1| R_{\pi/3} | e \rangle\\
\langle e| R_{\pi/3} | 1 \rangle & \langle e| R_{\pi/3} | e \rangle
\end{array}\right) = \frac{1}{2} \left( \begin{array}{cc}
1 & 1 \\
1 & 1
\end{array}\right). \label{eq:overlaps}
\end{equation}
Whereas the value of $\langle 1 | R_{\pi/3} | 1\rangle$ for the smaller torus is slightly above $1/2$ (see Fig.~\ref{fig:overlaps}(a)), repeating it for a bigger torus with $96$ sites (see Fig.~\ref{fig:overlaps}(b)), we agree with the prediction~\eqref{eq:overlaps}. For completeness, we also show $\langle 1 |e\rangle$: while the two MPS are orthogonal on the infinitely-long cylinder, it is a priori not guaranteed that they should remain orthogonal when making the MPS periodic on the torus \cite{Cincio13}. Hence, the fact that we find a small value $\langle 1 |e\rangle \approx 0.03$ confirms that the finite-size effects are rather small.

Another way of confirming that these two ground states correspond to the $1$ and $e$ anyon is by constructing the fixed-point wavefunctions, for which we find a large overlap. More precisely, we define $|1\rangle_\textrm{fix}$ as the state on the cylinder that corresponds to the superposition of all dimer configurations for which $\langle P \rangle_\textrm{loop} = \langle Q \rangle_\textrm{loop} = 1 $ around the circumference. The other three fixed-point wavefunctions $|e\rangle_\textrm{fix}, |m\rangle_\textrm{fix}, |f\rangle_\textrm{fix}$ are then obtained by respectively applying a $Q$, $P$ and $PQ$ string along the infinitely-long axis of the cylinder. We have confirmed that if we start from the fixed-point wavefunctions for $|1\rangle_\textrm{fix}$ and $|e\rangle_\textrm{fix}$ and perform imaginary time evolution, we converge toward the two ground states found by DMRG. This naturally gives us a way of also obtaining the ground states corresponding to the vison or magnetic particle $m$, and the fermion $f$. We have confirmed that the finite-size splitting of these four topological ground states decreases with circumference, plotted in Fig.~\ref{fig:gs} (for YC-4 and XC-8). Due to the inefficiency of imaginary time evolution compared to DMRG, we have not been able to prepare converged wavefunctions for $|m\rangle$ and $|f\rangle$ on YC-8, so we cannot consider their overlaps in Fig.~\ref{fig:overlaps}.

\begin{figure}[t]
    \centering
    \includegraphics[scale=1]{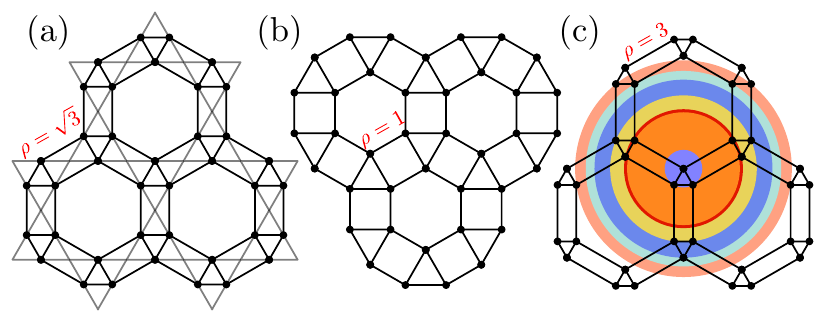}
    \caption{\textbf{The ruby lattice.} (a) Atoms on the \emph{links} of the kagome lattice form the \emph{vertices} of a ruby lattice where the rectangle has an aspect ratio $\rho = \sqrt{3}$. (b) The ruby lattice with $\rho=1$. (c) The ruby lattice with $\rho = 3$. The colored disks show seven distinct interaction distances; the phase diagram in Fig.~\ref{fig:phasediagram_aspect3} is obtained by including $V(r) = \Omega(R_b/r)^6$ for 16 distinct distances, coupling each site to $44$ other sites. }
    \label{fig:ruby}
\end{figure}

\begin{figure*}
    \centering
    \includegraphics[scale=0.34]{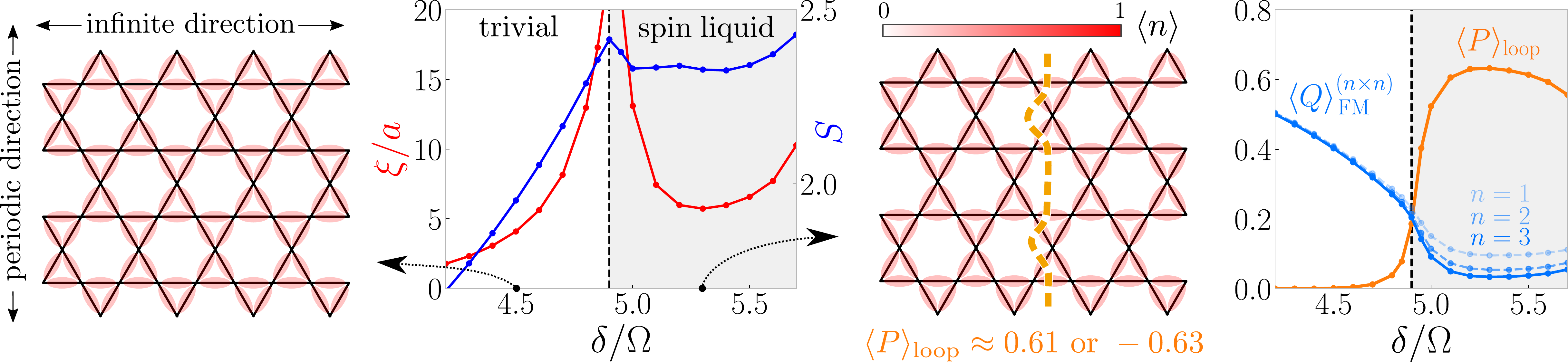}
    \caption{\textbf{Spin liquid on  ruby lattice ($\bm{ \rho=3 }$) with $\bm{ V(r)\sim 1/r^6 }$.} We consider the lattice in Fig.~\ref{fig:ruby}(c) for blockade radius $R_b = 3.8a$, keeping all interactions within a radius $r\leq 9a$ on the XC-8 geometry. There is a phase transition between two featureless phases, the latter having a large entanglement plateau. The spin liquid is characterized by the simultaneous vanishing of the off-diagonal string operator (i.e., the trivial phase is an $e$-condensate) and emergence of a large signal for the parity loop around the circumference. The latter also labels two of the degenerate ground states, as annotated on the density plot.
    The fact that this approximates a dimer model is evidenced by, e.g., $\langle n \rangle \approx 0.249$ for $\delta/\Omega=5.3$. To emphasize the connection to a dimer model, we plot the density on the links of the kagome lattice (i.e., $\rho=\sqrt{3}$). The correlation length is also expressed for this kagome geometry (i.e., the length of a triangle of the kagome lattice is $2a$).
    }
    \label{fig:phasediagram_aspect3}
\end{figure*}

A further characterization beyond topological order involves the implementation of symmetry, i.e., symmetry enrichment of topological order \cite{Wenbook}. This can be deduced from the relation to the kagome lattice dimer model, albeit in the absence of spin rotation symmetry (since monomers carry no spin). We expect the relevant projective symmetry group to be that of the bosonic mean field $Q_1=-Q_2$ state of Ref.~\cite{Sachdev_Triangle}, which has been related to other mean field representations in Refs.~\cite{Wang_2006, Lu_2017, LuRanLee}\footnote{The $Q_1=-Q_2$ state of  Ref.~\cite{Sachdev_Triangle} is equivalent to the [0Hex,0Rhom] of Ref.~\cite{Wang_2006}, which from Ref.~\cite{Lu_2017} is identified with the $Z_2[0,\pi]\beta$ fermionic state of Ref.~\cite{LuRanLee}.}. A caveat is that lattice symmetry enrichment, which implies a background `$e$' particle associated to each kagome site, can modify ground state overlap matrices for certain system sizes. 

\section{Prospects for realization and detection \label{sec:realization}}

In section~\ref{sec:blockade} we established that the Rydberg blockade model realizes a $\mathbb Z_2$ spin liquid for a range of parameters. The purpose of this section is twofold. First, we we would like to show that this result is not limited to the blockade model in Eq.~\eqref{eq:V}: the spin liquid persists on adopting the realistic Rydberg potential. Second, we would like to have a way to diagnose the existence of the spin liquid using probes available in Rydberg experiments. In light of that, we discuss how the string operators can be measured in the lab.

\subsection{Quantum liquid for $\sim 1/r^6$ potential and a family of ruby lattices \label{sec:vanderwaals}}

We now consider the Rydberg Hamiltonian in Eq.~\eqref{eq:ham} with the algebraically-decaying potential $V(r) = \frac{\Omega}{(r/R_b)^6}$; $R_b$ is commonly referred to as the (Rydberg) blockade radius due to sites well within this distance experiencing a large potential, effectively a blockade of the type discussed in section~\ref{sec:blockade}. Since $V(r)$ now explicitly depends on the distances between the atoms, it is important to discuss the geometry of the lattice. In the blockade model, we specified that the atoms live on the links of the kagome lattice (see Fig.~\ref{fig:model}(a)). These atoms form the vertices of the so-called ruby lattice, demonstrated in Fig.~\ref{fig:ruby}(a). In this particular case, we see that the rectangles of the ruby lattice have an aspect ratio $\rho = \sqrt{3}$. However, $\rho$ is a free tuning parameter\footnote{While only $\rho=1$ is an Archimedean lattice, the group of crystalline symmetries is the same for all $\rho$.}; as long as $\rho > 1/\sqrt{2} \approx 0.71$, the six sites nearest to a given site are the same set of points for which we defined the blockade in Fig.~\ref{fig:model}(a). If we thus choose $R_b$ to be large enough to enclose these six nearest sites (which are enclosed in a disk of radius $r_3/a = \sqrt{1+\rho^2}$), the resulting model approximates the blockade model. However, due to the $1/r^6$ interaction, we have additional longer-range couplings, and it is non-trivial to know whether or not the spin liquid will be stable to this. For this same reason, we will want to take $R_b$ smaller than the next interaction radius, i.e., as a rough guideline for where to search for the spin liquid:
\begin{equation}
\sqrt{1+\rho^2} < \frac{R_b}{a} < \min \left\{  \sqrt{3} \rho , \sqrt{1+ \sqrt{3}\rho + \rho^2} \right\}. \label{eq:Rb}
\end{equation}

\begin{figure}
    \centering
    \includegraphics[scale=0.28]{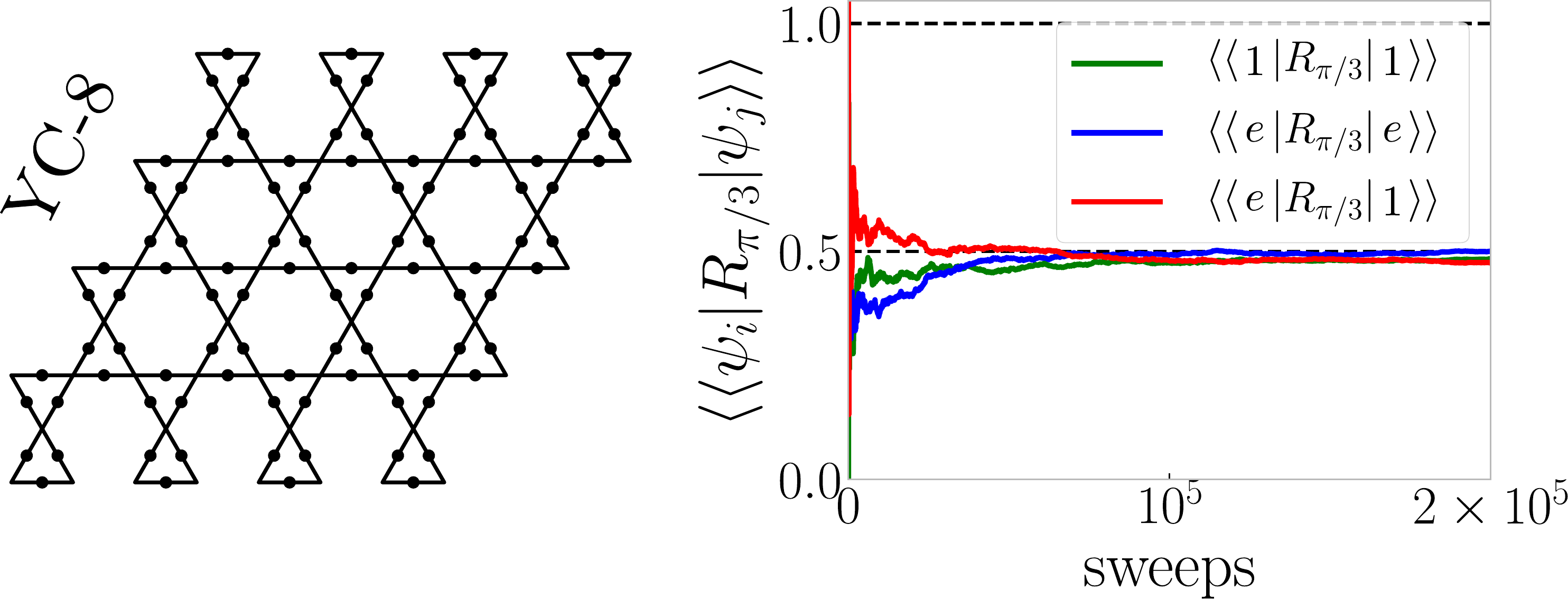}
    \caption{\textbf{Modular transformations on  ruby lattice ($\bm{ \rho=3 }$) with $\bm{ V(r)\sim 1/r^6 }$.} For blockade radius $R_b = 3.8a$ and detuning $\delta/\Omega = 5.5$, we consider two topologically distinct ground states on the YC-8 torus as shown (see main text and Appendix~\ref{app:torus} for details). The overlaps after a $\pi/3$-rotation agree with the universal value predicted for a $\mathbb Z_2$ spin liquid. (As in Fig.~\ref{fig:phasediagram_aspect3}, the simulation faithfully represents $V(r)$ within a distance $r \leq 9a$.)}
    \label{fig:overlaps_ruby}
\end{figure}

For concreteness, we consider the ruby lattice with $\rho=3$, depicted in Fig.~\ref{fig:ruby}(c). The rule of thumb in Eq.~\eqref{eq:Rb} suggests that we should look for the spin liquid in the range of $3.2 < R_b/a < 3.9$. We indeed find a spin liquid for the choice $R_b = 3.8a$ as shown in Fig.~\ref{fig:phasediagram_aspect3} for the XC-8 geometry. There are at least four independent indicators of the spin liquid: (i) we observe a phase transition between two featureless states of matter. We find that the latter is characterized by a large entanglement plateau, and its density is close to that of an ideal dimer state: for instance, at $\delta \approx 5.3 \Omega$, where the correlation length is minimal, we find $\langle n \rangle \approx 0.249$. (ii) Similar to the blockade model, the latter phase has a topological degeneracy: as shown in Fig.~\ref{fig:phasediagram_aspect3}, DMRG finds degenerate ground states with opposite signs for the parity loop around the circumference. (iii) The FM order parameter for the $Q$ string decays upon increasing the string length, signaling that we have exited the trivial (Higgs) phase (whereas absence of VBS order shows that we have not entered the confined phase). More precisely, we show $\langle Q\rangle_\textrm{FM}^{(n \times n)}$ where $n \times n$ counts the number of hexagons enclosed: see Fig.~\ref{fig:phasediagram_with_strings} for a sketch of $n=2$. (iv) We have also obtained $|1\rangle$ and $|e\rangle$ ground states on the YC-8 geometry. By putting these on a torus (see Section~\ref{sec:gs}) and properly orthogonalizing the resulting wavefunctions (see Appendix~\ref{app:torus}), we can calculate their overlaps after a $\pi/3$ rotation. As shown in Fig.~\ref{fig:overlaps_ruby}, these agree with the universal value $1/2$ predicted by the $S$ and $T$ matrices of $\mathbb Z_2$ topological order.

To numerically simulate the model with long-range interactions, we truncate $V(r) = \Omega(R_b/r)^6$ to zero beyond a distance $r> R_\textrm{trunc}$. The data for the XC-8 cylinder in Fig.~\ref{fig:phasediagram_aspect3} and the YC-8 cylinder in Fig.~\ref{fig:overlaps_ruby} is obtained for $R_\textrm{trunc}=9a$, which means that each site is coupled to 44 other sites (see Appendix~\ref{app:numeric} for details). We have moreover confirmed that the results are stable upon changing the interaction cutoff $R_\textrm{trunc}$, which was explicitly checked for $R_\textrm{trunc}=8a,9.5a,10a$.
As a note of caution, let us mention that if $R_\textrm{trunc}$ is small, the physics can depend on it. E.g., if only neighboring triangles are coupled (for $\rho=3$ this corresponds to $R_\textrm{trunc} = (3+\sqrt{3})a$, indicated by the dark blue disk in Fig.~\ref{fig:ruby}(c)), we indeed find a spin liquid phase with $R_b =3.8a$ as above. However, upon including one further interaction radius, the spin liquid is destabilized. This result can be understood intuitively by noting that this additional coupling punishes hexagon flipping resonances which are essential for a spin liquid. Including yet more interactions again induces a spin liquid, eventually in a stable way as mentioned above. Note that since an XC-4 cylinder only has a circumference $L_\textrm{circ} = (\sqrt{3}+3\rho)a \approx 10.7a$, one should not compare entanglement entropies between XC-8 and XC-4 since we must always ensure that $R_\textrm{trunc} \leq L_\textrm{circ}/2$. This explains why---unlike for the blockade model---we did not discuss topological entanglement entropy for this long-range interacting model.

The above establishes our main goal of showing the presence of a $\mathbb Z_2$ spin liquid for a model with Van der Waals interactions. Note that the model has multiple tuning parameters that could further stabilize this topological phase: the lattice aspect ratio $\rho$, the Rydberg blockade radius $R_b$ and the detuning $\delta/\Omega$. It would be interesting to use this freedom to find the global minimum of the correlation length in the spin liquid phase. We leave such an exhaustive search through this three-parameter phase diagram to future work.
For the case of the ruby lattice with $\rho=\sqrt{3}$ (corresponding to atoms living on the links of the kagome lattice),
we find a spin liquid for $R_b \approx 2.4a$ upon including the first four interaction distances. However, we see indications that further-range interactions tend to destabilize the spin liquid at $\rho = \sqrt{3}$, unlike in the case reported with $\rho=3$. A detailed examination of the case $\rho = \sqrt{3}$ will appear in forthcoming work.

Let us also briefly note that while our numerical results are for the cylinder geometry, an experimental realization would of course have open boundary conditions. The main difference is that then there are no topologically non-trivial loops (i.e., all loops are contractible) and correspondingly the ground state is unique. Nevertheless, a topological ground state degeneracy can be recovered by either puncturing the system, or by considering mixed boundary conditions. Both mechanisms are explained in detail in Section~\ref{sec:qubit}, where we also consider numerical results for the strip geometry.

\subsection{Measuring an off-diagonal string by transforming it into a diagonal string \label{sec:offdiagonal}}

In section~\ref{sec:strings} we introduced the two topological string operators associated to the $\mathbb Z_2$ lattice gauge theory. These can be very useful for identifying the spin liquid and its nearby phases (see Fig.~\ref{fig:phasediagram_with_strings}). Fortunately, the parity string $P$ can be straightforwardly measured in the lab since it is diagonal in the occupation basis and can be read off from the snapshots of the Rydberg states. The off-diagonal string $Q$ is more challenging to measure directly. We now show that by time-evolving with a quenched Rydberg Hamiltonian, it becomes a diagonal observable, making it experimentally accessible. Aside from its practical significance, this result is also conceptually valuable since it gives a concrete duality transformation between the two strings. Due to the local constraint, such a duality is rather non-trivial.

\begin{figure}
    \centering
    \includegraphics[scale=1]{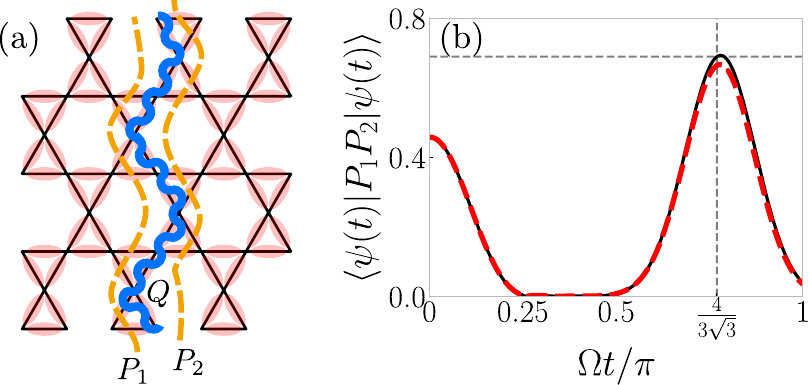}
    \caption{\textbf{Measuring the off-diagonal string operator through a quench protocol.} (a) The vertical direction is periodic on the cylinder. The off-diagonal string $Q$ (blue wiggly line) can be obtained by measuring the diagonal string $P_1P_2$ (orange dashed lines) after a time-evolution with a nearest-neighbor Rydberg blockade Hamiltonian to time $\Omega t = \frac{4\pi}{3\sqrt{3}}$ (see Eq.~\eqref{eq:rotation}). (b) Starting from the spin liquid ground state of the Rydberg Hamiltonian with $R_b= 3.8a$ and $\delta/\Omega = 5.3$ on a ruby lattice with aspect ratio $\rho=3$ (see Fig.~\ref{fig:phasediagram_aspect3}), we measure $\langle P_1P_2 \rangle$ after time-evolving with either the nearest-neighbor Rydberg blockade model (solid black line) or the ground state Hamiltonian quenched to $R_b = 2$ (red dashed line). The horizontal gray dashed line denotes the ground state value for $\langle Q \rangle$. \color{black} }
    \label{fig:time_evol}
\end{figure}

To implement this rotation, we consider the Rydberg Hamiltonian at zero detuning with a complex phase factor in the Rabi oscillation\footnote{This can be engineered by combining the original Hamiltonian with an appropriately-timed evolution where the detuning is dominant, i.e., using $e^{-i\alpha \sigma^z/2} \sigma^x e^{i\alpha \sigma^z/2} = \cos(\alpha) \sigma^x + i \sin(\alpha) \sigma^y$.}:
\begin{equation} H' = \frac{\Omega}{2} \sum_{\bm i} \left( i e^{i\alpha} b_{\bm i}^\dagger +h.c.\right)  + \frac{1}{2}\sum_{\bm i,\bm j} V(|\bm i - \bm j|) \; n_{\bm i} n_{\bm j}.  \label{eq:ham2}
\end{equation}
The essential idea is to consider the evolution under a Rydberg blockade localized on individual triangles of the ruby lattice, i.e., $V(r_1) = +\infty$ and $V(r)=0$ otherwise (see Fig.~\ref{fig:model}(a) for the definition of $r_1$). 

Since the blockade now only acts within triangles of the ruby lattice, time-evolving with the above Hamiltonian amounts to an on-site unitary transformation. It is thus sufficient to consider a single triangle, and by writing the $P$ and $Q$ operators defined in Fig.~\ref{fig:strings}(a) as $4\times 4$-matrices acting on the Hilbert space of a single triangle, one straightforwardly derives (see Appendix~\ref{app:rotate} for details):
\begin{equation}
\begin{tikzpicture}[baseline=-0.65ex]
\node at (0,0) {\includegraphics[scale=0.29]{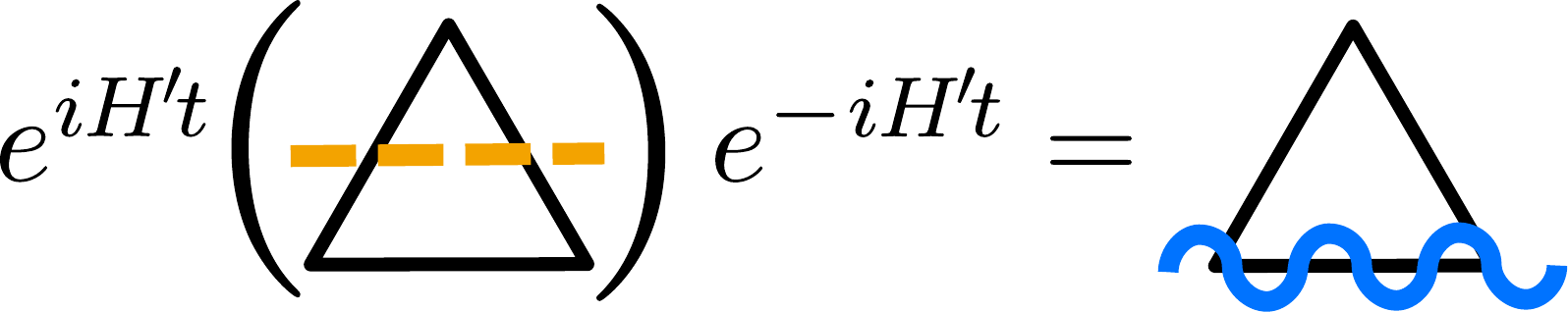}};
\node at (3.7,0) {for $\Omega t = \frac{4\pi}{3\sqrt{3}}$.};
\end{tikzpicture} \label{eq:rotation}
\end{equation}
Thus, one can effectively measure $Q$ along a string by first time-evolving with $H'$ and then measuring the $P$ string on the resulting state.

If the aspect ratio $\rho$ of the ruby lattice is not too close to unity, one can approximate this nearest-neighor blockade Hamiltonian by quenching $R_b$ in between the first two radii, i.e., $1 < R_b/a < \rho$. For instance, we have confirmed that for $\rho=3$, a quench from $R_b = 3.8a$ (where we found the spin liquid in Fig.~\ref{fig:phasediagram_aspect3}) to $R_b = 2a$ gives virtually indistinguishable results from time-evolving with the nearest-neighbor blockade (see Fig.~\ref{fig:time_evol}). In either case, we confirm that the value of the diagonal correlator at $\Omega t= \tau_\textrm{dual} := \frac{4\pi}{3\sqrt{3}} \approx 2.42$ correctly reproduces the ground state expectation value for the off-diagonal string operator\footnote{Note that the $Q$-loop in Fig.~\ref{fig:time_evol} is dual to \emph{two} parity strings. This is because it has triangles on both side of the string. Other loops with triangles on only one side---such as a loop around a hexagon---will be dual to a single parity string.}.
In the experimental set-up, the blockade radius $R_b$ can be effectively tuned by changing $\Omega$. In particular, since $V(r) = \Omega (R_b/r)^6$, reducing $R_b$ from $3.8a$ to $2a$ (as in the above example) corresponds to changing $\Omega$ by a factor $(3.8/2)^6 \approx 47$. While appreciable, this factor is achievable with current methods.

\begin{figure}
    \centering
    \includegraphics[scale=0.36]{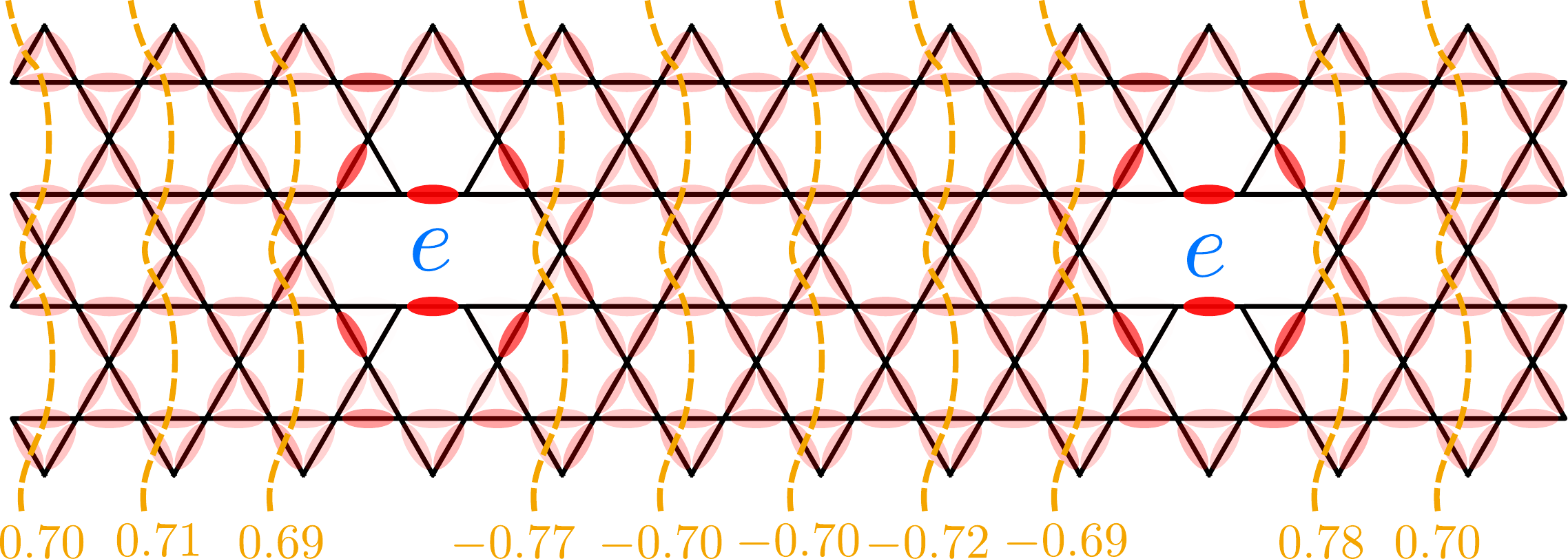}
    \caption{\textbf{A trapping potential for $e$-anyons.} The ground state (here on a cylinder) for a lattice where four sites around a vertex have been removed captures an $e$-anyon. This can be read off from the expectation value of the parity loops (dashed orange lines) around the circumference: if two neighboring loops have opposite sign, then a charge is enclosed. \label{fig:etrap}}
\end{figure}

\begin{figure*}
    \centering
    \includegraphics[scale=1]{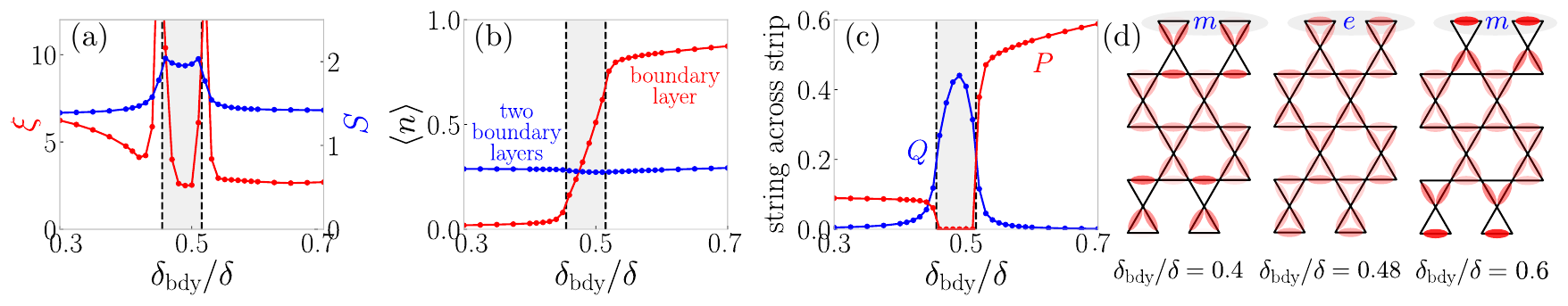}
    \caption{\textbf{Boundary phase diagram of the blockade model.} We consider an infinitely long strip of the XC-8 geometry: the bulk is the spin liquid at $\delta/\Omega=1.7$, but we tune $\delta$ on the outermost boundary links. (a) The correlation length diverges at two boundary phase transitions; in the intermediate shaded regime, the entanglement is increased. (b) The small and large $\delta_{\textrm{bdy}}$ phases have a classical-like dimer filling at the boundary, whereas the intermediate regime has a compressible boundary. (c) By calculating the string operators from boundary-to-boundary, we diagnose the small and large (intermediate) $\delta_{\textrm{bdy}}$ phases as having $m$-condensed ($e$-condensed) boundaries. (d) Density plots $\langle n \rangle$ in the three boundary regimes. The strip is infinitely long (finite) in the horizontal (vertical) direction.}
    \label{fig:bdyphasediagram}
\end{figure*}

\section{Towards fault-tolerant quantum memory \label{sec:qubit}}
Part of the reason that topologically ordered phases of matter are of great interest is that they can serve as a means of potentially creating fault-tolerant quantum memories based on degenerate topological ground states \cite{Kitaev_2003}. We have already encountered such degeneracies associated to a $\mathbb Z_2$ spin liquid in section~\ref{sec:gs}. However, this example utilized periodic boundary conditions, which is not natural in an experimental setting. Fortunately, topologically-distinct ground states can also arise for systems with boundaries. This can occur both for systems with punctures/holes (which one can interpret as a sort of boundary), as well as systems with mixed boundary conditions. Either of these options requires the knowledge of how to realize distinct topological boundary conditions. Another important ingredient is the trapping of anyons whose braiding implements gates on the quantum bits. We first analyse these two ingredients, after which we discuss what one can do with them.

\subsection{Trapping an $e$-anyon \label{sec:e}}

If one wishes to braid with anyons, one has to be able to localize them to a particular region. Since the $e$-anyon in this model corresponds to a monomer (e.g., see the discussion in section~\ref{sec:dimer}), a natural way of trapping it is by forcing a certain vertex to have no dimer touching it. This can be done by either simply removing the atoms on these bonds, or by lowering the detuning $\delta$. We numerically confirm that this works: Fig.~\ref{fig:etrap} shows the result of removing two such vertices on XC-8 for the blockade model at $\delta/\Omega =1.7$. Since parity loops measure the charge enclosed in a given loop, the nonzero charge localized on these defects can be inferred from comparing the sign of the parity loops along the cylinder. In fact, we even see that the two $e$-anyons are connected by a gauge string where the parity loops are negative.

Note that the actual removal of atoms is not required: the same effect is obtained by locally setting the detuning $\delta \ll  -|\Omega|$. By adiabatically changing the detuning, this anyon can potentially be moved around at will, allowing for control over an $e$-anyon. Similar approaches can potentially be explored to trap and control $m$-anyons as well. Even in the absence of such an $m$-anyon, the $e$-anyon can already be used for non-trivial braiding, as we will discuss in section~\ref{sec:surface}.

\subsection{Boundary phase diagram \label{sec:boundary}}

There are two topologically-distinct boundary conditions for a $\mathbb Z_2$ spin liquid. 
These are characterized by whether the $e$ or $m$ anyon condenses at the edge. It is no coincidence that the trivial and VBS phase are also described as condensates (see section~\ref{sec:strings}): if one interprets a boundary as a spatial interface from the topological phase to a non-topological phase, it is natural that the characterization of the nearby phases carries over to describe boundary conditions. Similarly, these $e$ and $m$ condensates along the boundary can be diagnosed using the string operators introduced in section~\ref{sec:strings}. More precisely, $m$-boundaries ($e$-boundaries) have long-range order for the $P$-string ($Q$-string).

Simply terminating the lattice---keeping all the Hamiltonian terms that fit on the remaining geometry---will tend to stabilize the $m$-boundary. Indeed, since boundary dimers experience less repulsion, they will prefer to arrange in a classical pattern with few fluctuations, giving long-range order to the diagonal string operator $P$. To stabilize the $e$-boundary condition, we need to enhance such boundary fluctuations. One way of doing so is by changing the detuning $\delta$ along the boundary sites, searching for the sweet spot where the dimers are suspended between the two classical (empty or filled) configurations.

We numerically determine the resulting boundary phase diagram for the blockade model on an infinitely-long strip geometry, where we choose the bulk to be deep in the spin liquid at $\delta/\Omega=1.7$. The results are shown in Fig.~\ref{fig:bdyphasediagram}. In line with the above expectation, we see that before we change the boundary detuning, i.e., $\delta_\textrm{bdy} = \delta$, the strip realizes an $m$-boundary as evidenced by the large response for the end-to-end parity string. As we reach $\delta_\textrm{bdy} \approx 0.5 \delta$, there is a boundary phase transition (where the correlation length diverges along the infinite direction) after which the parity string dies out, making way for a strong signal for the $Q$ string. In this regime, we stabilize the $e$-boundary. As we further decrease $\delta_\textrm{bdy} \to 0$, we are effectively removing these links from the model, with the remaining geometry again spontaneously realizing an $m$-boundary. This picture is also confirmed by the density plots and the $\langle n\rangle$ curve: it is only in the intermediate regime---corresponding to the $e$-boundary---that the edge dimers are fluctuating.

\subsection{Topological degeneracy on the plane \label{sec:surface}}

\begin{figure}
    \centering
    \includegraphics[scale=1]{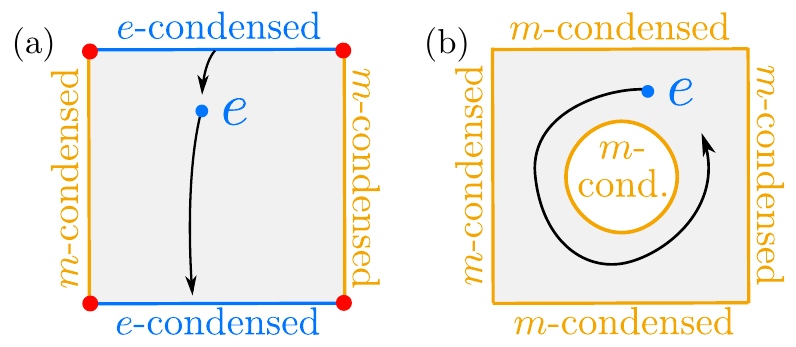}
    \caption{\textbf{Topological degeneracy in planar geometry.} (a) Alternating $e$- and $m$-condensed boundaries imply a twofold degeneracy. One way of understanding this is in terms of the Majorana zero modes (red dots) that live at the points where the boundary condition changes \cite{Bombin10}; due to the global emergent fermion parity having to be unity, these four Majorana modes only give rise to a twofold degeneracy. If we label states using the $P$-string connecting the left and right boundaries (see main text), then pulling an $e$-anyon out of one $e$-condensed boundary to another effectively toggles the states in this two-level system. (b) An annulus geometry with $m$-condensed boundaries also has a twofold degeneracy. Moving the $e$-anyon around the hole will toggle the states. Since $e$-anyons can only be created in pairs, there will be another $e$-anyon which we do not move (not shown).}
    \label{fig:planar}
\end{figure}

With the knowledge of the above boundary phase diagram, it is now straightforward to construct a rectangular geometry with a topological ground state degeneracy. A schematic picture is shown in Fig.~\ref{fig:planar}(a): a square slab where the four boundaries are alternatingly $e$- and $m$-condensed. One way of understanding this twofold degeneracy is as follows: one can imagine extracting a single $e$-anyon from the top boundary (after all, it is an $e$-condensate), dragging it through the deconfined bulk, and depositing it at the bottom boundary. Similarly, one can do the same for an $m$-anyon from left to right. Due to the mutual statistics of $e$ and $m$, these two processes anti-commute, implying a degeneracy.

Let us now address how to physically label this two-level system, or equivalently, how to read out a given state. If the spin liquid was in a fixed-point limit---similar to the toric code \cite{Kitaev_2003}---then the topological string operators $P$ and $Q$ (defined in section~\ref{sec:strings}) would be exact symmetries of the model. I.e., the logical $\sigma^z_\textrm{logic}$ ($\sigma^x_\textrm{logic}$) operator could then be identified with any $P$-($Q$-)string connecting the $m$-condensed ($e$-condensed) boundaries. However, our system is not at a fixed-point limit, such that acting with these $P$ and $Q$ string operators need not stay with this subspace; relatedly, we cannot label our system in terms of eigenstates of $P$ or $Q$. Fortunately, using the idea of the FM order parameter encountered in section~\ref{fig:strings}, we can define properly-normalized expectation values:
\begin{equation}
\begin{tikzpicture}[baseline=-0.65ex,scale=0.23]
\node at (0,0) {\includegraphics[scale=0.17]{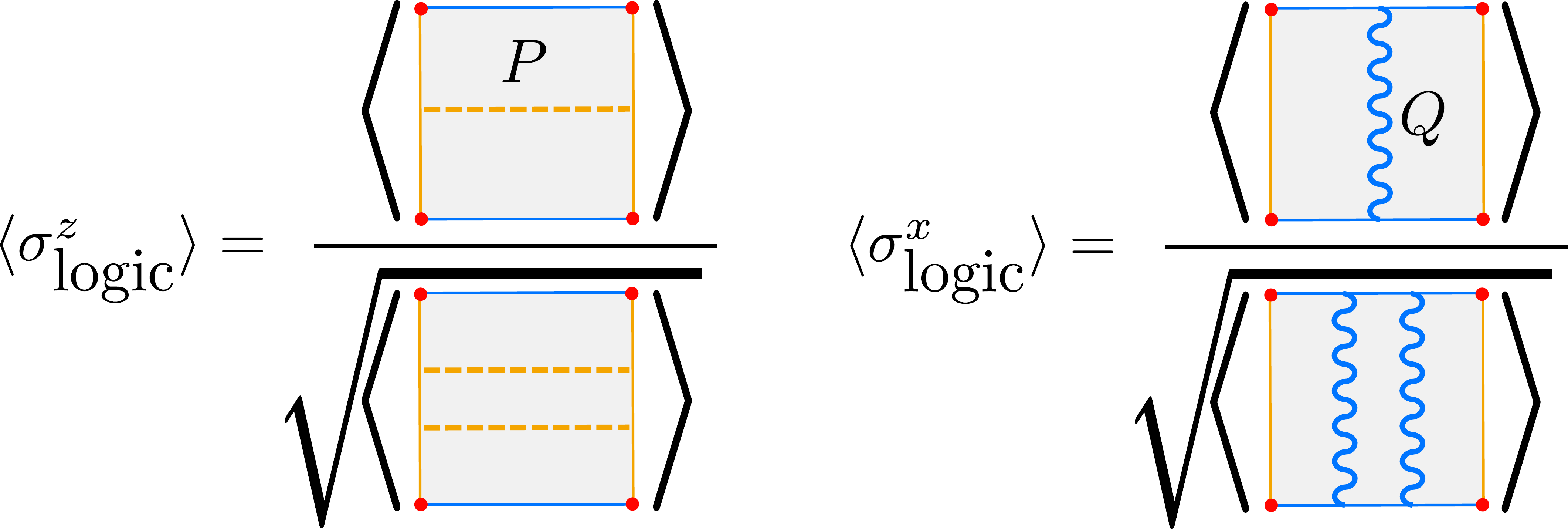}};
\end{tikzpicture}
\label{eq:logical}
\end{equation}
It is worth pointing out that unlike the numerators in Eq.~\eqref{eq:logical}, the denominators do not depend on the logical state of the system\footnote{To see this, remember that the degeneracy could be interpreted as being a consequence of moving $m$- or $e$-anyons between the corresponding condensed boundaries, but these commute with \emph{pairs} of topological string operators.} and hence they only need to be determined once for any particular architecture.

\begin{figure}
    \centering
    \includegraphics[scale=0.35]{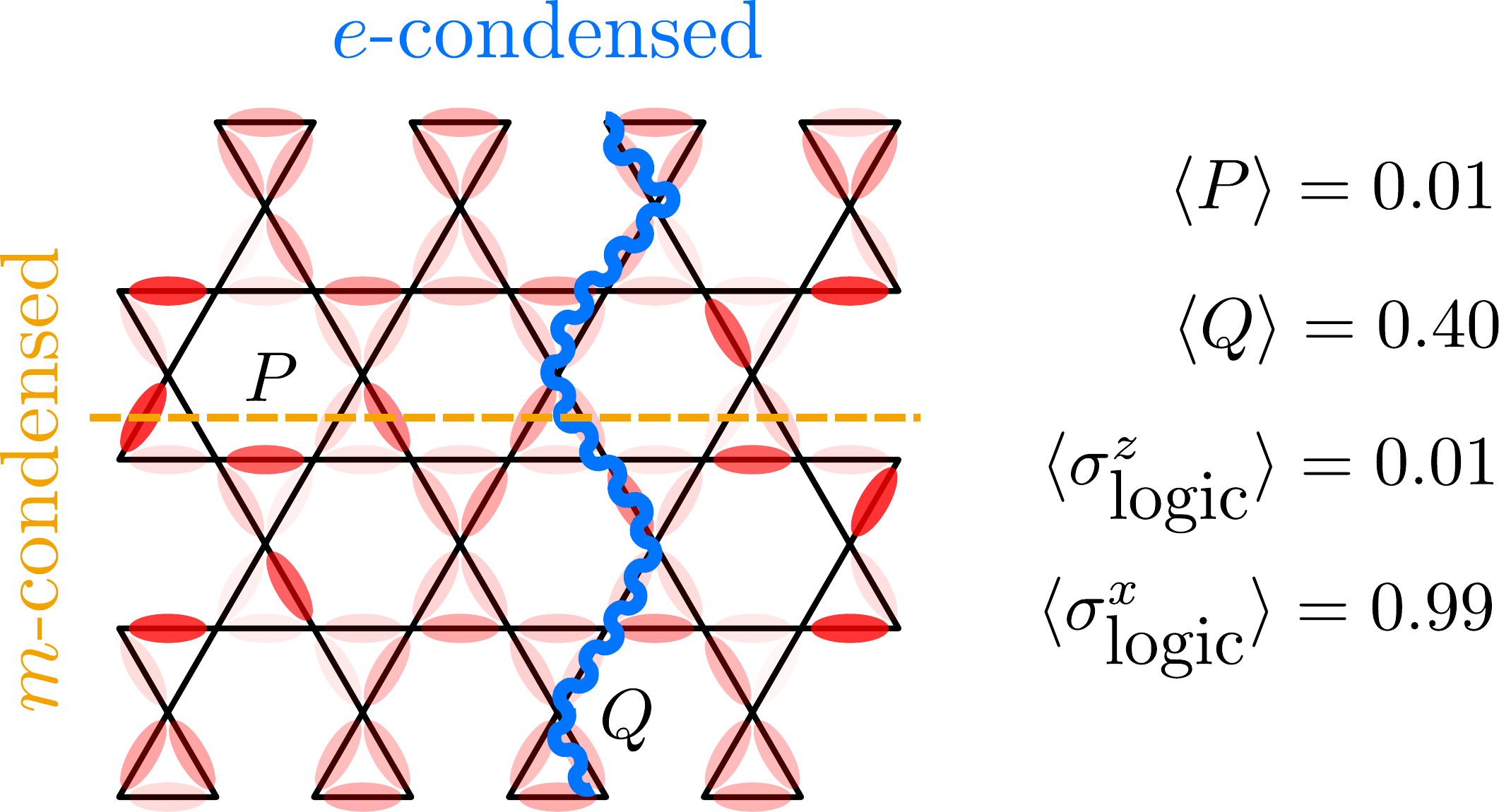}
    \caption{\textbf{Read-out of a topological ground state.} We consider the blockade model for $\delta/\Omega = 1.7$ on a finite sample with open boundaries, as shown. Moreover, we change the laser detuning on the top and bottom boundary to $\delta_\textrm{bdy} = 0.48\delta$. From the boundary phase diagram in Fig.~\ref{fig:bdyphasediagram}, we know that this realizes the $e$-condensed boundary, whereas the left and right boundaries are $m$-condensates. For the ground state of this system, we show the values for the two type of topological string operators which connect their corresponding condensates. Upon using the FM normalization (see Eq.~\eqref{eq:logical}), the read-out for the logical variables gives a state that lies along the $x$-axis of the Bloch sphere. Note that both string operators can be experimentally measured using the prescription in section~\ref{sec:offdiagonal}.}
    \label{fig:qubit}
\end{figure}

To illustrate that this procedure is meaningful and well-defined, let us consider a simulated example, as shown in Fig.~\ref{fig:qubit}. The top and bottom boundaries were tuned to be $e$-condensed using the boundary phase diagram in Fig.~\ref{fig:bdyphasediagram}, setting $\delta_\textrm{bdy}=0.48\delta$. First, we observe that $\langle Q \rangle \neq 0$ when it connects the top and bottom boundaries; this is consistent with these being $e$-condensed. Moreover, we see that $\langle P \rangle \approx 0$ from left-to-right. This suggests that this state lies entirely along the logical-$x$ axis (in the Bloch sphere picture). To confirm that $\langle P \rangle \approx 0$ is not due to an error in the boundary conditions (after all, the same result would arise for a parity string connecting two $e$-condensed boundaries), we confirm that for two parallel parity strings connecting the two $m$-condensed boundaries, we obtain the nonzero response $|\langle P_1 P_2 \rangle| \approx 0.46$. As an additional sanity check, we confirmed that this same double-parity-string gives a zero response when running from top-to-bottom. Finally, using the FM-prescription in Eq.~\eqref{eq:logical}, we obtain that the logical state indeed lies along the $x$-axis: $\langle \sigma^x_\textrm{logic}\rangle \approx 1$.

We thus have a way of labeling and reading out our topological quantum state. Let us now consider the question of \emph{initialization}. We work with the logical basis $\{ |0\rangle, |1\rangle \}$, defined by $\langle n | \sigma^z_\textrm{logic} |n \rangle = (-1)^n$. We can create $|0\rangle$ by starting with a sample which only has an $m$-condensed boundary---such that the parity string is a fixed positive value---and then adiabatically create an $e$-condensed boundary as follows:
\begin{center}
\includegraphics[scale=0.26]{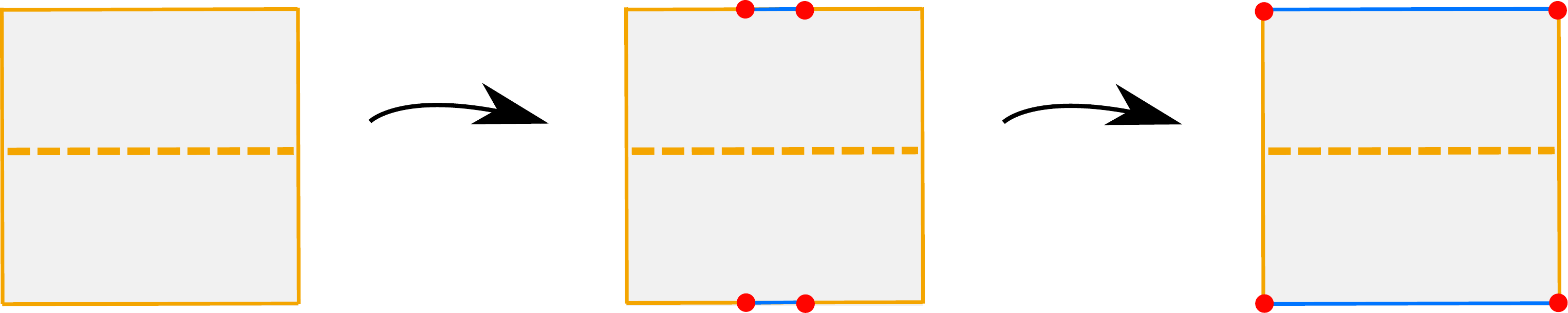}
\end{center}
In the above sequence, we also show the parity string whose value will not change throughout this process, such that we arrive at $|0\rangle$. To initialize into $|1\rangle$, we can now use the fact that we know how to pin an $e$-anyon (see section~\ref{sec:e}). We can thus dynamically change the detuning to pull an $e$-anyon out the top $e$-condensed boundary and move it into the bottom $e$-condensed boundary, as sketched in Fig.~\ref{fig:planar}(a). This implements the logical $\sigma^x_\textrm{logic}$ gate, mapping $|0\rangle \to |1\rangle$.

One can repeat the above steps for the alternative architecture of an annulus, shown in Fig.~\ref{fig:planar}(b). In particular, in this case the logical state is toggled by braiding the $e$-anyon around the $m$-condensed hole. More generally, one can create multiple $e$- and $m$-condensed holes in a given sample. Braiding these (by dynamically changing the parameters of the Hamiltonian) potentially gives another handle on topological processing of quantum information \cite{Bravyi98,Cong17}.

\section{Outlook}

We have demonstrated that Rydberg blockade on the ruby lattice can be utilized to stabilize a $\mathbb Z_2$ spin liquid. The underlying mechanism is that of a monomer-dimer model where single-site monomer fluctuations induce the dimer resonances necessary for a resonating valence bond state. This same picture also leads to a specific form of the two topological string operators. The spin liquid---stable to longer-range $V(r)\sim 1/r^6$ interactions---can be characterized by these string observables in experiment where they are measurable by appealing to a dynamic protocol. 
Moreover, we showed that this system could be used to explore topological 
quantum memories by localizing anyons, realizing conjugate boundary conditions which create degeneracy on the plane, and reading out quantum states. We note that given the detailed differences between our platform and the exact toric code model, these implementations required new insights. While the robustness of these techniques in the presence of realistic imperfections (such as, e.g., spontaneous emission) will need to be carefully explored, it is important to emphasize that the atom array platform offers fundamentally new tools for probing and manipulating topological quantum matter.  

Specifically, the theoretical predictions outlined above can be probed using programmable quantum simulators based on neutral atom arrays. In particular, the required atom 
arrangements can be realized using demonstrated atom sorting techniques, while relevant effective blockade range can be readily implemented using laser excitation into Rydberg states with large principle quantum number $60<n<100$.
Note that in designing the appropriate atom arrays, careful choice of  atomic separations and Rydberg states should be made to avoid molecular resonances \cite{Derevianko15} that could modify the blockade constraint.
The spin liquid phase can be created via adiabatic sweep of laser detuning, starting from the disordered phase to a desired value of positive detuning, as demonstrated previously for one-dimensional \cite{Bernien17,Keesling19} and two-dimensional \cite{Ebadi20,Scholl20} systems. For typical parameters, corresponding to effective Rabi frequencies in the range of few MHz, such adiabatic sweeps can be carried out with minimal decoherence in systems potentially exceeding 200 atoms. We note that the topologically ordered state is separated from the trivial product state by a single continuous transition which is favorable for preparation. A number of tools can be deployed to identify and study the transition into spin liquid state that lacks local order parameter. While the transition point can be identified by  measuring filling fraction (see Fig.~\ref{fig:n_and_P}), much more detailed investigations can be carried out by measuring the expectation value  of  parity operators (Fig.~\ref{fig:phasediagram_with_strings}) associated with various loops. Remarkably, both $P$  and $Q$ operators can be efficiently measured, by either directly analysing the signal shot images or carrying out this analysis following qubit rotation in the dimer basis associated with individual triangles (as explained in section~\ref{sec:offdiagonal}). The latter can be realized using resonant atomic driving with appropriately chosen parameters. 
Moreover, the topological entanglement entropy can be potentially obtained by measuring the second Renyi entropy \cite{Ekert02,Islam15,Kaufman16,Linke18,Brydges19} for different regions, as described in Refs.~\cite{Kitaev06b,Levin06}. Together with control over boundaries and exploration of samples with non-trivial topology, these methods constitute a unique opportunity for detailed explorations of spin liquid  states with accuracy and sophistication  not accessible with any 
other existing approaches. 

Furthermore, this work opens up a number of very intriguing avenues that can be explored in the framework introduced here. These range from exploration of non-equilibrium dynamical properties of spin liquid states in response to rapid changes of various Hamiltonian parameters, to experimental realization and detection of anyons with non-trivial statistics. In particular, anyon braiding can be explored by using time-varying local potentials (see e.g. section~\ref{sec:e}).
Moreover, approaches to improve the stability of TQL and realization of more  exotic spin liquid states can potentially be realized by additional engineering of interaction potentials, using e.g. long-lived hyperfine atomic states \cite{Zeiher16,Levine19,MSS20b}. In particular, approaches involving optical lattice \cite{Zeiher16} and Rydberg dressing \cite{MSS20} could be explored to realize a broader variety of spin liquid states.
Finally, we note that the blockade model is essentially an Ising model on the ruby lattice. Such models could be implemented in other ways, e.g., in arrays of superconducting qubits \cite{Harris10,Hauke20,Kjaergaard20}, magnets with strongly-anisotropic exchange \cite{Coldea10}, or perhaps even in recently developed two-dimensional materials \cite{Moire}. Potentially, these systems can be used for the realization of topologically-protected quantum bits, with an eye towards developing new, robust approaches to manipulating quantum information.

{\em Note Added:} An independent work which will appear in this same posting also studies the quantum phases of Rydberg atoms but in a different arrangement, where atoms occupy {\it sites} of the kagome lattice \cite{Rhine_unpublished}.
\begin{acknowledgments}
We thank Dave Aasen, Marcus Bintz, Soonwon Choi, Sepehr Ebadi, Xun Gao, Marcin Kalinowski, Alexander Keesling, Harry Levine, Hannes Pichler, Subir Sachdev, Giulia Semeghini, Norm Yao and Mike Zaletel for useful discussions. The DMRG simulations were performed using the Tensor Network Python (TeNPy) package developed by Johannes Hauschild and Frank Pollmann \cite{Hauschild18}. This work was supported by the Harvard Quantum Initiative Postdoctoral Fellowship in Science and Engineering (R.V.) and by the Simons Collaboration on Ultra-Quantum Matter, which is a grant from the Simons Foundation (651440, A.V.).
M.D.L. was supported by the U.S. Department of Energy (Grant DE-SC0021013),
the Harvard-MIT Center for Ultracold Atoms (Grant PHY-1734011), 
the Army Research Office (Grant W911NF2010082),
and the National Science Foundation (Grant PHY-2012023).
The computations in this paper were run on the FASRC Cannon and Odyssey clusters supported by the FAS Division of Science Research Computing Group at Harvard University.
\end{acknowledgments}
\bibliography{main.bbl}

\onecolumngrid
\appendix
\setcounter{figure}{0}
\renewcommand\thefigure{\thesection.\arabic{figure}}

\section{Numerical details \label{app:numeric}}

As mentioned in Section~\ref{sec:blockade}, in this work we consider two types of cylinders of the kagome lattice, called XC or YC. This naming convention was introduced by Ref.~\cite{Yan_2011}: if one considers the kagome lattice as depicted in Fig.~\ref{fig:phasediagram} then the XC cylinder has its infinite direction along the $x$-axis, whereas for the YC cylinder this is along the $y$-axis (and in both cases, the `C' simply stands for `cylinder'). As a consequence, we see that the finite periodic direction of the YC cylinder runs along of the bonds of the kagome lattice.

The DMRG simulations were performed using the open-access Tensor Network Python (TeNPy) package developed by Johannes Hauschild and Frank Pollmann \cite{Hauschild18}, version $0.7.2$. Although DMRG is a method for one-dimensional systems, it can be used for cylinder geometries by snaking through the system (i.e., giving all sites a one-dimensional labeling) \cite{Stoudenmire12}. The cost one pays for this is that couplings which used to be nearby in the two-dimensional geometry will typically become further-range couplings in this effective one-dimensional labeling. To obtain the ground state, we start with a low bond dimension, say $\chi=100$ or $\chi=200$, and repeat DMRG for successively larger values of $\chi$ until physical observables were no longer found to change. For most plots in this work, $\chi=1000$ is sufficient, although in certain cases we have gone up to $\chi=2000$. As an additional sanity check that the bond dimension was chosen large enough to accurately encode the ground state physics, it is very useful to consider the density $\langle n \rangle$ on sites which are equivalent on the cylinder but \emph{not} equivalent in the effective one-dimensional labeling: if $\chi$ is too low, their expectation values will typically not coincide; it is only when the ground state correctly converges to a ground state on the two-dimensional cylinder that the densities on such sites will coincide. This is thus a very powerful indicator of convergence.

For systems on an infinitely-long cylinder, we used a translation-invariant ansatz consisting of a certain number of rings.
If this number is chosen too small to fit a particular VBS pattern, this issue shows up in an inability of DMRG to converge to a stable state (and sometimes it leads to a large norm error due to the tendency to form a cat state).
In such cases, the number of independent rings was increased until the state converged. This is how we found the VBS phase in Fig.~\ref{fig:phasediagram}. Sometimes this VBS phase can get stuck in a local minimum: for instance, we also found ground states where the two pinwheels in Fig.~\ref{fig:phasediagram} had opposite orientations. For this reason, we started in a variety of distinct initial states, and we found that the global minimum occurered for the VBS pattern shown in Fig.~\ref{fig:phasediagram}. When the phase was trivial or a spin liquid phase, we found that an ansatz of a single ring was sufficient to obtain a converged state (although we confirmed that the result was unchanged upon increasing the number of rings), except for the YC-6 geometry (for which the entanglement entropy appears in Fig.~\ref{fig:Stopo}) which has a Lieb-Schultz-Mattis anomaly---there a two-ring ansatz was necessary, even in the trivial and spin liquid phases. Let us also mention that the correlation length $\xi$ is obtained via the standard MPS procedure: one diagonalizes the transfer matrix---if its largest eigenvalue is normalized to be unity, then the absolute value of its second largest eigenvalue is $e^{-1/\xi}$.

\begin{figure}
    \centering
    \includegraphics[scale=0.7]{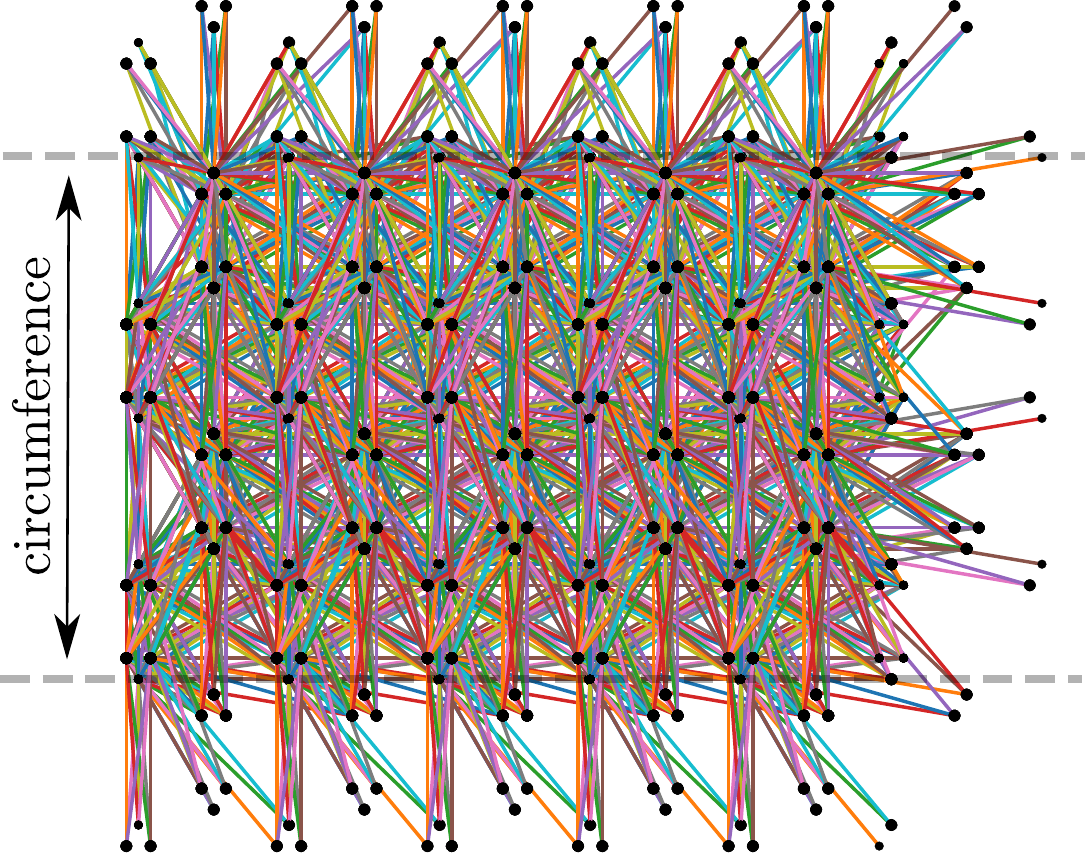}
    \caption{\textbf{Connectivity graph for Van der Waals interactions on the ruby lattice.} Black dots denote the ruby lattice with $\rho=3$ (also see Fig.~\ref{fig:ruby}(c)). Each line represents a coupling in $V(r) = \Omega(R_b/r)^6$ that is included in the numerics for the phase diagram in Fig.~\ref{fig:phasediagram_aspect3}. The gray dashed lines denote how this is wrapped into an XC-8 cylinder; any site of the ruby lattice outside this region can be identified with a site inside this region. The cylinder is infinitely-long in the horizontal direction.}
    \label{fig:couplings}
\end{figure}

In Section~\ref{sec:realization}, we considered a model on the ruby lattice with long-range Van der Waals interactions. In particular, for the ruby lattice with $\rho=3$ and blockade radius $R_b = 3.8a$, the data in Figs.~\ref{fig:phasediagram_aspect3} and \ref{fig:overlaps_ruby} was obtained for $V(r) = \Omega(R_b/r)^6$ for $r\leq 9a$ and $V(r) = 0$ for $r> 9a$. Its connectivity graph is shown in Fig.~\ref{fig:couplings} for the XC-8 geometry.

\setcounter{figure}{0}

\section{Scaling of Fredenhagen-Marcu order parameter \label{app:scaling}}

\begin{figure}
    \centering
    \color{black}
    \begin{tikzpicture}
    \node at (0,0){\includegraphics[scale=0.44]{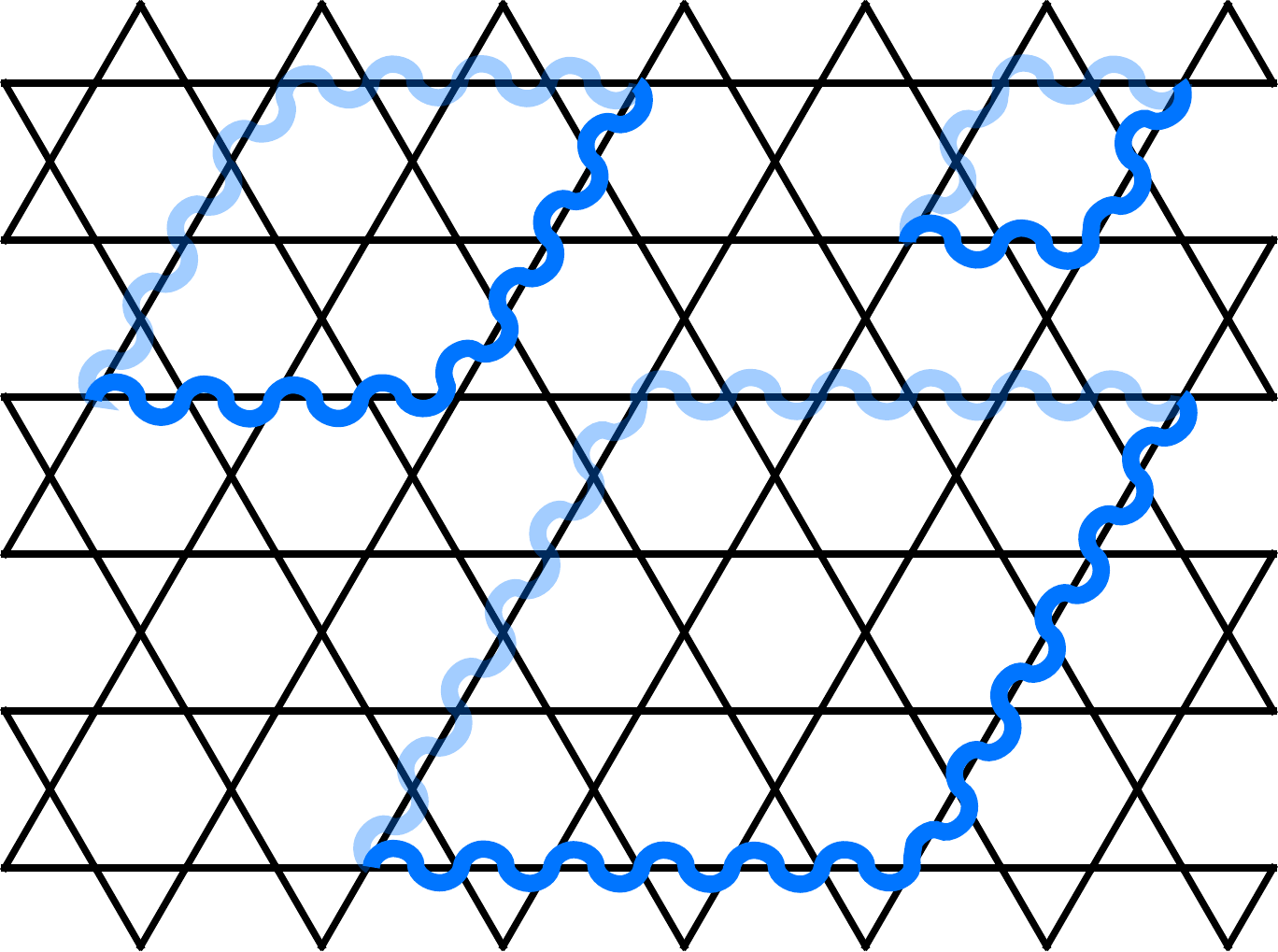}};
    \node at (8.2,-0.2){\includegraphics[scale=0.42]{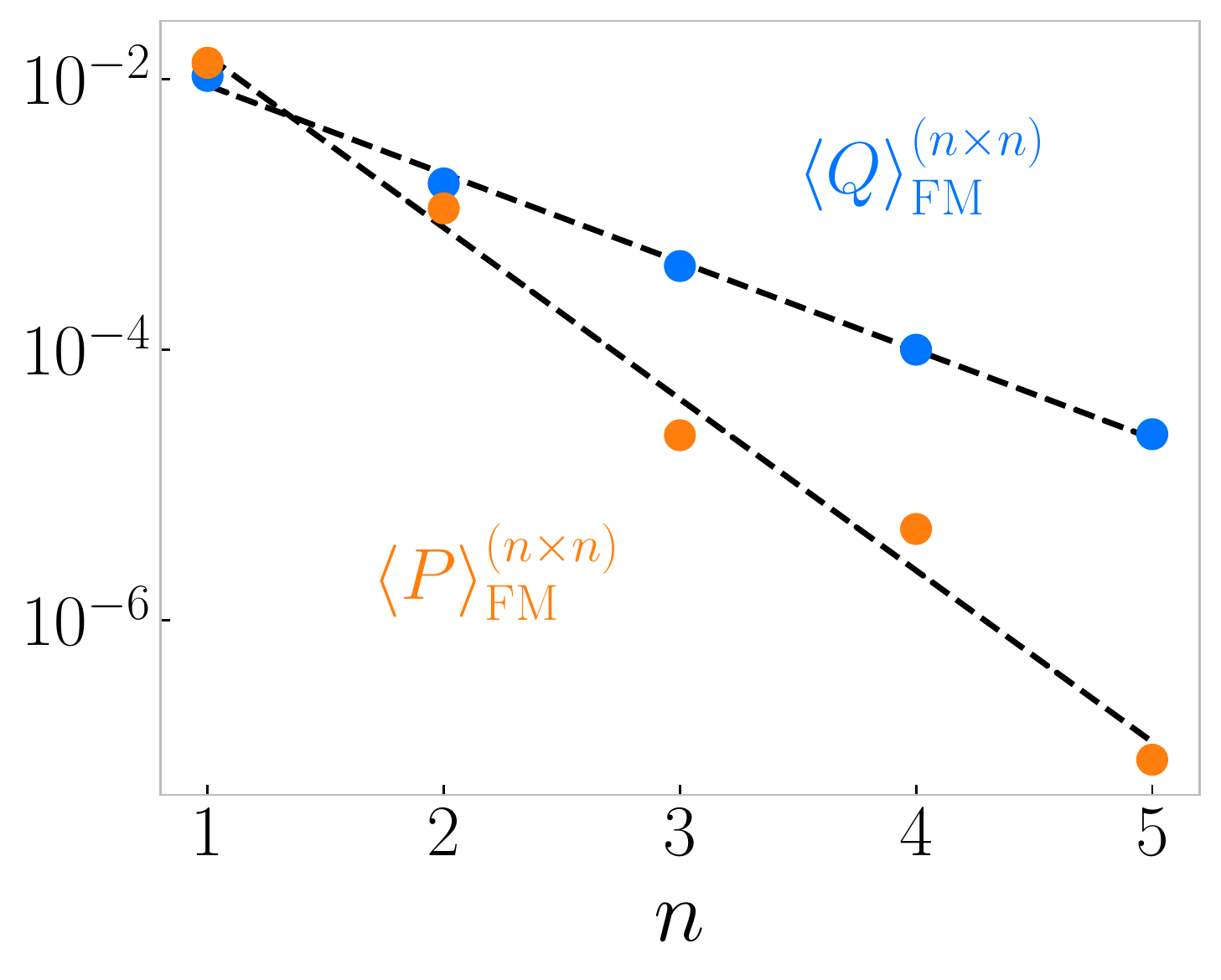}};
    \node at (-3.7,2) {(a)};
    \node at (4.8,2) {(b)};
    \end{tikzpicture}
    \caption{\textbf{The FM string order parameter in the blockade model.} (a) The lattice shown is the XC-12 cylinder (periodic along vertical direction, infinite along horizontal direction). The blue lines denote the $Q$-strings defining the FM order parameter $\langle Q \rangle_\textrm{FM}^{(n\times n)}$ ($n=1,2,3$), which is normalized by the square root of the closed string (see Fig.~\ref{fig:strings}(c) for the general definition). (b) Values for the FM order parameters in the blockade model on XC-12 deep in the spin liquid, $\delta/\Omega = 1.7$. Both strings decay to zero exponentially with the length of the string, a property that is unique to the deconfined phase.}
    \label{fig:scaling}
\end{figure}

In Fig.~\ref{fig:phasediagram_with_strings} of the main text, we show the FM string order parameters $\langle Q\rangle_\textrm{FM}^{(n\times n)}$ and $\langle P\rangle_\textrm{FM}^{(n\times n)}$ for $n=2$ (sketches of the string geometry are also shown in that figure). The plot suggests that these strings decay to zero in the intermediate phase, consistent with this being the deconfined phase. To confirm this claim, here we go deep in the spin liquid, $\delta/\Omega = 1.7$, and scale the FM string order parameters with their length $n$. For clarity, we sketch the strings that define $\langle Q\rangle_\textrm{FM}^{(n\times n)}$ in Fig.~\ref{fig:scaling}(a) for $n=1,2,3$. The values for $n=1,2,3,4,5$ (for both types of strings) are shown in panel Fig.~\ref{fig:scaling}(b) which was obtained on the XC-12 geometry with bond dimension $\chi=1400$: we see that these values decay to zero exponentially with the length of the string, as expected in the deconfined phase \cite{Marcu86}.

\begin{figure}
    \centering
    \color{black}
    \begin{tikzpicture}
    \node at (-0.7,0){\includegraphics[scale=0.44]{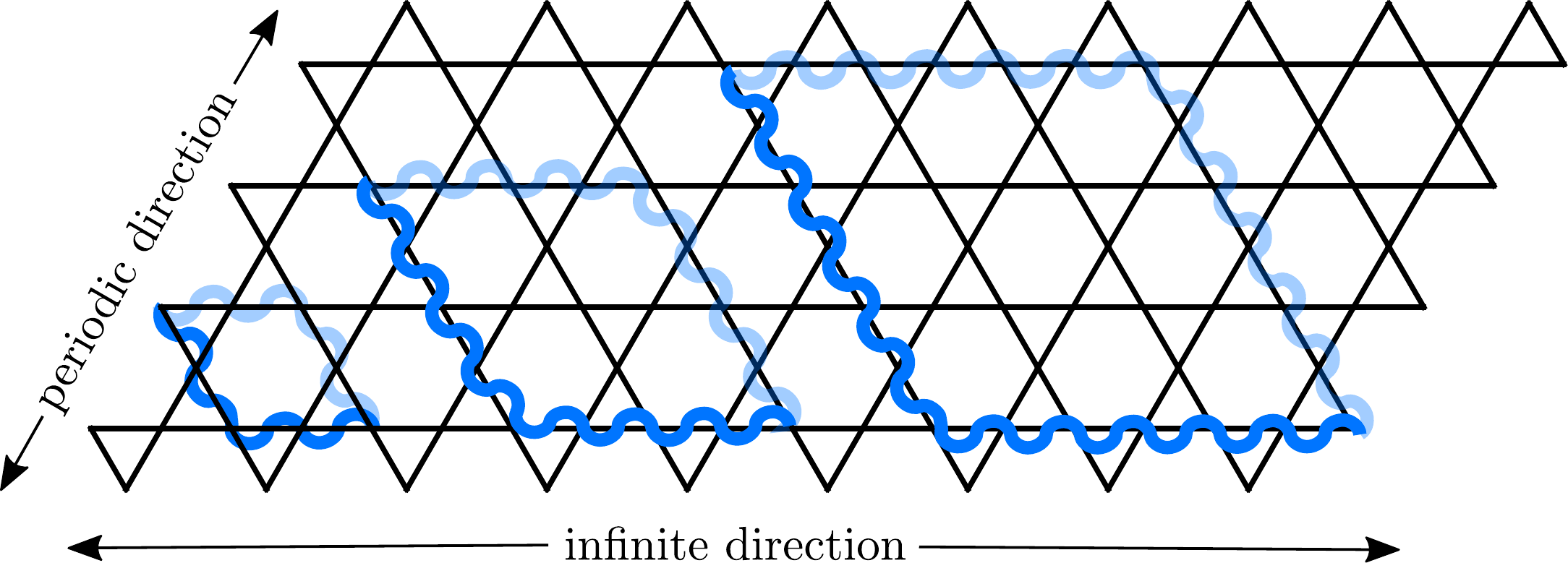}};
    \node at (8.2,-0.2){\includegraphics[scale=0.38]{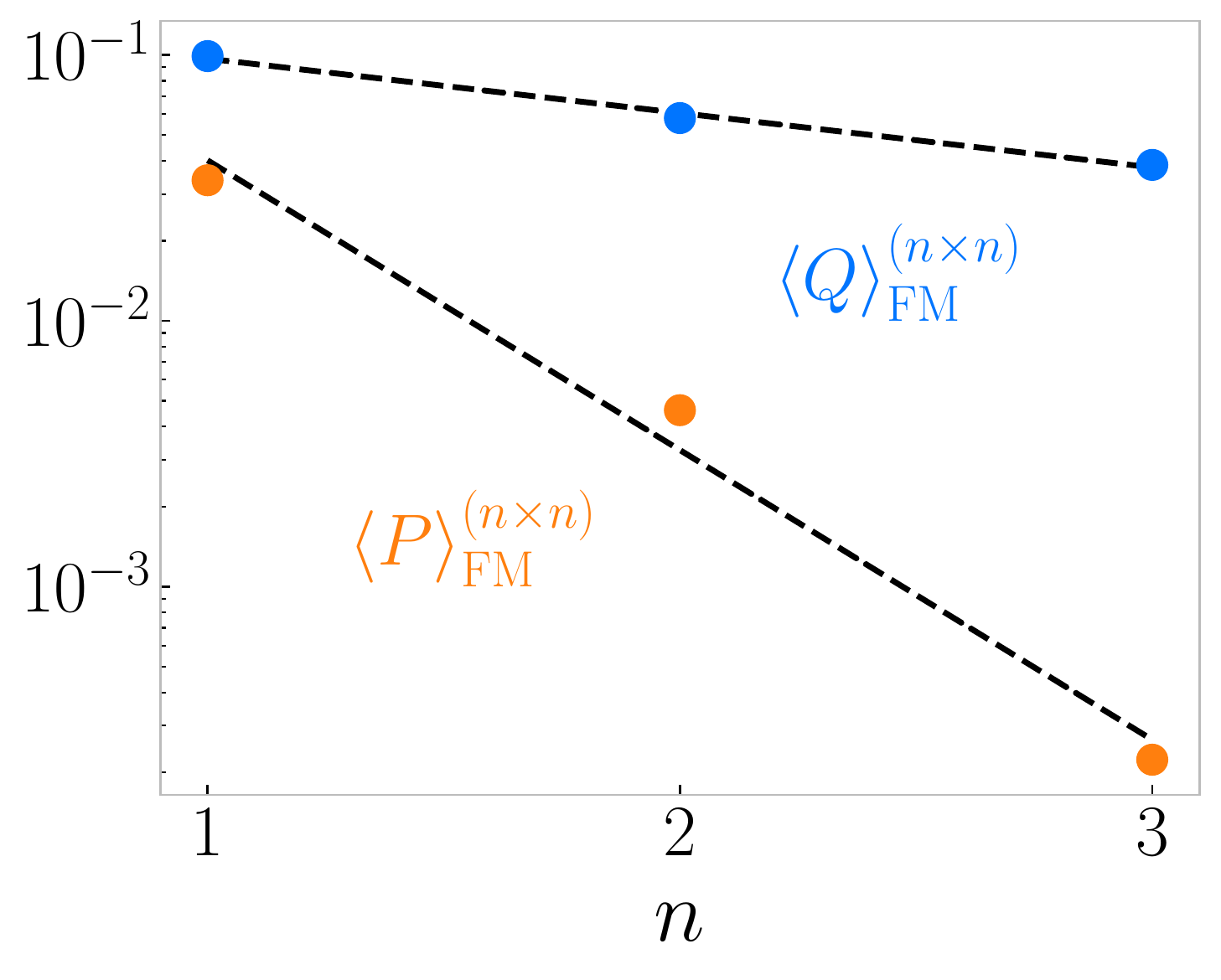}};
    \node at (-5.2,1.6) {(a)};
    \node at (4.95,1.6) {(b)};
    \end{tikzpicture}
    \caption{\textbf{The FM string order parameter in the $V(r) \sim 1/r^6$ model.} Similar to Fig.~\ref{fig:scaling}, except: (a) we work on the YC-8 cylinder; (b) the DMRG results for the FM order parameters are now for the ruby lattice model discussed in Section~\ref{sec:vanderwaals}, in particular: the blockade radius is $R_b = 3.8a$, the detuning $\delta/\Omega=5.5$, and we faithfully represent $V(r) = \Omega(R_b/r)^6$ within a distance $r \leq 9a$.}
    \label{fig:scaling_ruby}
\end{figure}

For the spin liquid in the model with $V(r) \sim 1/r^6$ interactions (see Section~\ref{sec:vanderwaals}), we cannot go up to XC-12 cylinders. We are thus limited in repeating the same analysis, but for completeness, we present the results on the largest cylinder accessible in this case: YC-8, shown in Fig.~\ref{fig:scaling_ruby}(a). Although Fig.~\ref{fig:scaling_ruby}(b) only presents results for three distinct FM string sizes, the qualitative behavior is consistent with that of a spin liquid, and is similar to what we observed in Fig.~\ref{fig:scaling}(b). Indeed, both the $P$ and $Q$-FM strings decay as a function of string length, with the former decaying faster. To give a more quantitative comparison, let us note that the dashed lines in Fig.~\ref{fig:scaling}(b) give $\exp(-a \times n)$ with $a\approx 1.5$ for $Q$ and $a \approx 2.9$ for $P$, whereas in Fig.~\ref{fig:scaling_ruby}(b) we obtained $a \approx 0.5$ for $Q$ and $a \approx 2.5$ for $P$. Hence, the results for $P$-strings are similar, whereas the decay of the $Q$-string is three times steeper in the blockade model. This is consistent with the phase diagrams in Fig.~\ref{fig:phasediagram} and Fig.~\ref{fig:phasediagram_aspect3}, where we observed the blockade model is closer to a fixed-point model (i.e., it has a smaller correlation length).

\section{Topological ground states on the torus \label{app:torus}}

As explained in great detail in Ref.~\cite{Cincio13}, ground states on a torus geometry can be approximated by first using DMRG to obtain the ground state on an infinitely-long cylinder and then simply evaluating the resulting matrix product state wavefunction on the torus (by identifying the appropriate virtual indices). In Ref.~\cite{Cincio13} this was moreover used to construct minimally entangled states (MES) on the torus: more precisely, Ref.~\cite{Jiang12} had clarified that the topologically distinct ground states found by DMRG on an infinitely-long cylinder are naturally MES, and if finite-size effects are small, using the prescription of Ref.~\cite{Cincio13}, this gives MES on the torus.

However, this need not be true if finite-size effects are strong enough to induce virtual anyon loops that wind around the torus. For concreteness, let us denote the direction along the circumference of cylinder as `vertical' and the infinite direction along the cylinder axis as `horizontal'. Upon putting this wavefunction on a torus (i.e., the horizontal direction is made finite and periodic), then virtual anyonic fluctuations could wind around the horizontal direction and connect distinct topological sectors. This mean that the resulting state is no longer a MES.

To make this more precise, it is useful to characterize MES as states which are eigenstates of the topological line operators along the vertical direction. Let us denote $P_\textrm{ver}$ and $Q_\textrm{ver}$ as the loop operators around this vertical direction; similarly $P_\textrm{hor}$ and $Q_\textrm{hor}$ denote loops around the finite horizontal direction of the torus. We would like to obtain the MES $|1\rrangle$ and $|e\rrangle$ which are characterized (in the idealized case) by eigenvalues $Q_\textrm{vert}= +1$ and $P_\textrm{vert} = \pm 1$; this also means that while they are not eigenstates of the horizontal loops, they would have a vanishing expectation value, e.g., $\llangle 1|Q_{\textrm{hor}}|1 \rrangle = 0$ (since $P_\textrm{vert} Q_\textrm{hor} = - Q_\textrm{hor} P_\textrm{vert}$). If we denote the states obtained from placing the cylinder ground states (with the same vertical loop observables) on the torus geometry as $|1\rangle$ and $|e\rangle$, then these do not automatically coincide with the aforementioned $|1\rrangle$ and $|e\rrangle$ states: finite-size fluctuations can induce a nonzero value for the horizontal strings, e.g., $\langle 1 |Q_\textrm{hor} | 1\rangle \neq 0$. In the context of the present work, the dominant fluctuations are in the $e$-anyons (see also Fig.~\ref{fig:scaling}(b)). Indeed, phenomenologically we find that $\langle P_\textrm{hor} \rangle \approx 0$ for our torus ground states, both for the blockade and Van der Waals model. The fluctuations that induce $\langle Q_\textrm{hor} \rangle \neq 0$ are closely linked to $\langle 1 | e\rangle \neq 0$: indeed, the $|1\rangle$ and $|e\rangle$ ground states are related by acting with a $Q$-string along the horizontal direction.

In the blockade model, we find that $\langle 1 |e \rangle \approx 0$ (see Fig.~\ref{fig:overlaps}), such that to a good approximation, $|1\rrangle \approx |1\rangle$ and $|e\rrangle \approx |e \rangle$. However, for the models with Van der Waals interactions considered in Section~\ref{sec:realization}, we find $\langle 1 | e\rangle \approx 0.4$ on the YC-8 torus of the ruby lattice with $\rho=3$, blockade radius $R_b = 3.8 a$ and detuning $\delta/\Omega = 5.5$. In this case, we thus need to work some more to obtain a good approximation for $|1 \rrangle$ and $|e\rrangle$. Indeed, since $Q$-loops are more strongly fluctuating, it is better to consider the superpositions $|1\rangle \pm |e\rangle$. On the infinitely-long cylinder, these are eigenstates of $Q_\textrm{hor}$ and $Q_\textrm{ver}$. Since fluctuations in $P$-loops are found to be negligible, an accurate identification on the torus is $\frac{ |1 \rrangle \pm |e \rrangle}{\sqrt{2}} \approx  \frac{ |1\rangle \pm |e\rangle }{\sqrt{2} Z_\pm}$. There is a proportionality factor $Z_\pm$ since $\frac{ |1\rangle \pm |e\rangle }{\sqrt{2}}$ are not properly normalized: their norm is $\frac{ \langle 1 | 1 \rangle + \langle e | e \rangle}{2} \pm \langle 1 | e \rangle \approx 1 \pm \langle 1 | e \rangle$ (where we used the realness condition of the wavefunction), i.e., $Z_\pm = \sqrt{1\pm \langle 1 |e\rangle} $.

In conclusion, we have:
\begin{equation}
\begin{array}{ccc}
|1\rrangle &= &\left( \alpha_+ + \alpha_-\right) |1\rangle + \left( \alpha_+ -\alpha_-\right) |e\rangle \\
|e\rrangle &= &\left( \alpha_+ - \alpha_-\right) |1\rangle + \left( \alpha_+ + \alpha_-\right) |e\rangle
\end{array} \qquad \textrm{with } \alpha_\pm = \frac{1}{2\sqrt{1\pm\langle 1 |e \rangle}}.
\end{equation}
It is for these MES that we plot the overlaps after $\pi/3$-rotation in Fig.~\ref{fig:overlaps_ruby}, finding excellent agreement with the prediction for $\mathbb Z_2$ topological order \cite{Zhang12}.

\section{Duality between topological string operators \label{app:rotate}}

\setcounter{figure}{0}

Here we prove Eq.~\eqref{eq:rotation}. For this, let us first label the four basis states in a single triangle as follows:
\begin{center}
\includegraphics[scale=0.35]{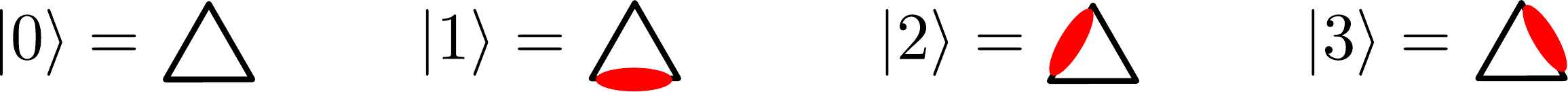}
\end{center}
Then the $P$ and $Q$ string operators (defined in Fig.~\ref{fig:strings}(a)) can be written as:
\begin{equation}
\begin{tikzpicture}[baseline=-0.65ex]
\node at (0,0){\includegraphics[scale=0.35]{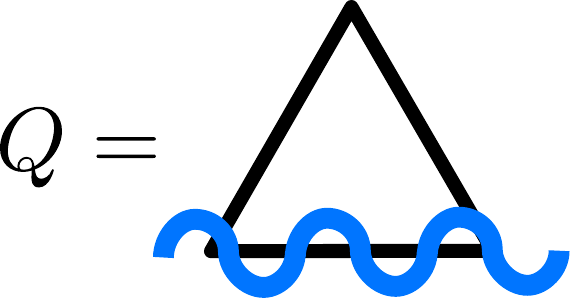}};
\end{tikzpicture} = \left(  \begin{array}{cccc} 0 & q & 0 & 0 \\
q^* & 0 & 0 & 0\\
0 & 0 & 0 & 1 \\
0 & 0 & 1 & 0 \end{array} \right) \qquad
\textrm{and} \qquad
\begin{tikzpicture}[baseline=-0.65ex]
\node at (0,0){\includegraphics[scale=0.35]{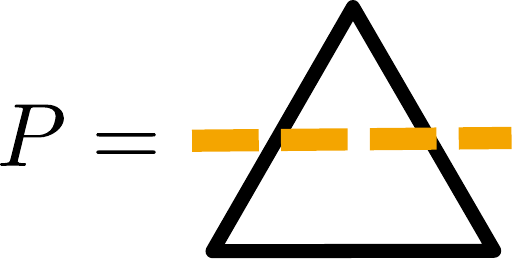}};
\end{tikzpicture} = \left(  \begin{array}{cccc} 1 &0 & 0 & 0 \\
0 & 1 & 0 & 0\\
0 & 0 & -1 & 0 \\
0 & 0 & 0 & -1 \end{array} \right),
\end{equation}
where we introduced $q = e^{-i\alpha}$.

The Hamiltonian defined in Eq.~\eqref{eq:ham2} does not couple distinct triangles, so it is sufficient to prove the claim for a single triangle. Then Eq.~\eqref{eq:ham2} becomes
\begin{equation}
H' = \frac{\Omega}{2} \sum_{i \in \triangle} P(i q^* b_i^\dagger - i q b)P = \frac{ i\Omega }{2} \left(  \begin{array}{cccc}  0 & -q & -q & -q \\
q^* & 0 & 0 & 0\\
q^* & 0 & 0 & 0 \\
q^* & 0 & 0 & 0 \end{array} \right)
= \frac{\Omega}{2} \times V D V^\dagger
\end{equation}
where
\begin{equation}
D = \sqrt{3} \left(  \begin{array}{cccc} 1 &0 & 0 & 0 \\
0 & -1 & 0 & 0\\
0 & 0 & 0 & 0 \\
0 & 0 & 0 & 0 \end{array} \right) \qquad \textrm{and} \qquad V =  \frac{1}{\sqrt{6}} \left(  \begin{array}{cccc} -iq\sqrt{3} & iq\sqrt{3} & 0 & 0 \\
1 & 1 & -2 & 0 \\
1 & 1 & 1 & -\sqrt{3} \\
1 & 1 & 1 & \sqrt{3} \end{array} \right).
\end{equation}

The time-evolution operator is thus
\begin{equation}
t = \frac{2}{\Omega} \times \frac{2\pi}{3} \times \frac{1}{\sqrt{3}} \qquad \Rightarrow \qquad  e^{-iH't} = V \left(  \begin{array}{cccc} e^{-2\pi i /3} &0 & 0 & 0 \\
0 &e^{2\pi i /3} & 0 & 0\\
0 & 0 & 1 & 0 \\
0 & 0 & 0 & 1 \end{array} \right) V^\dagger
= - \frac{1}{2} \left(  \begin{array}{cccc}1 & q & q & q \\
-q^* & -1 & 1 & 1\\
-q^* & 1 & -1 & 1 \\
-q^* & 1 & 1 & -1\end{array} \right).
\end{equation}

Then
\begin{equation}
e^{iHt} \left(  \begin{array}{cccc} 1 &0 & 0 & 0 \\
0 & 1 & 0 & 0\\
0 & 0 & -1 & 0 \\
0 & 0 & 0 & -1 \end{array} \right) e^{-iHt} = \left(  \begin{array}{cccc} 0 & q & 0 & 0 \\
q^* & 0 & 0 & 0\\
0 & 0 & 0 & 1 \\
0 & 0 & 1 & 0 \end{array} \right) .
\end{equation}

Q.E.D.

\end{document}